\documentclass{article}
\usepackage[utf8]{inputenc}
\usepackage{subfiles}
\usepackage{amsmath}
\usepackage{subfigure}
\usepackage{epsfig}
\usepackage{soul}
\usepackage{float}
\usepackage{setspace}
\usepackage[ruled,vlined]{algorithm2e}
\usepackage{tabularx,booktabs}
\usepackage{graphicx}
\usepackage{makecell}
\usepackage{dcolumn}
\usepackage{array}
\usepackage{collcell}
\usepackage{comment}
\usepackage{url}
\usepackage[utf8]{inputenc}
\usepackage[normalem]{ulem}
\usepackage{changepage,threeparttable} % for wide tables
\usepackage{xcolor}
\usepackage[most]{tcolorbox}
\definecolor{myred}{rgb}{0.958, 0, 0}

\definecolor{red}{rgb}{0, 0, 0}
\definecolor{browna}{rgb}{0.76,0.72,0.65}
\definecolor{brownb}{rgb}{0.71,0.69,0.65}

\newcolumntype{H}{>{\setbox0=\hbox\bgroup}c<{\egroup}@{}}
\useunder{\uline}{\ul}{}

\title{Open Source Software Sustainability: Combining Institutional Analysis and Socio-Technical Networks}

\author{Likang Yin\thanks{Corresponding Author}\\ Computer Science Department, UC Davis
\and Mahasweta Chakraborti\\ Communication Department, UC Davis
\and Charles Schweik\\ Environmental Conservation Department, UMass Amherst
\and Seth Fray\\ Communication Department, UC Davis
\and Vladimir Filkov\thanks{Corresponding Author}\\ Computer Science Department, UC Davis}

\begin{document}
\maketitle

\section*{Abstract}
Open Source Software (OSS) forms much of the fabric of our digital society, especially successful and sustainable ones. But many OSS projects do not become sustainable, resulting in abandonment and even risks for the world's digital infrastructure. 
Prior work has looked at the reasons for this mainly from two very different perspectives. 
In software engineering, the focus has been on understanding success and sustainability from the socio-technical perspective: the OSS programmers' day-to-day activities and the artifacts they create. 
In institutional analysis, on the other hand, emphasis has been on institutional designs (e.g., policies, rules, and norms) that structure governance.
Even though each is necessary for a comprehensive understanding of OSS projects, the connection and interaction between the two approaches have been barely explored.

In this paper, we make the first effort toward understanding OSS project sustainability using a dual-view analysis, by combining institutional analysis with socio-technical systems analysis.
In particular, we
(i) use linguistic approaches to extract institutional rules and norms from OSS contributors' communications to represent the evolution of their governance systems, and
(ii) construct socio-technical networks based on longitudinal collaboration records to represent each project's organizational structure.
We combined the two methods and applied them to a dataset of developer traces from 253 nascent OSS projects within the Apache Software Foundation (ASF) incubator.
We find that the socio-technical and institutional features relate to each other, and provide complementary views into the progress of the ASF's OSS projects.
Refining these combined analyses can help provide a more precise understanding of the synchronization between the evolution of institutional governance and organizational structure. 

\section{Introduction}

%\documentclass[main.tex]{subfiles}
%\begin{document}

Open Source Software (OSS) is a multi-billion dollar industry. 
A majority of modern businesses, including all major tech companies, rely on OSS without even knowing it.
OSS contributions are an important manifestation of computer-supported collaborative work, for the high degree of technical literacy typical of OSS contributors. 
Even though this popularity attracts many software developers to open source, more than 80\% of OSS projects are abandoned~\cite{schweik2012internet}.

The failure of collaborative work in OSS has received attention from two perspectives. In software engineering, the focus has been on understanding success and sustainability from the socio-technical perspective: the OSS developers' day-to-day activities and the artifacts they create. 
In the management policy domain, on the other hand, emphasis has been on institutional designs (e.g., policies, rules, and norms) that structure governance \textcolor{red}{and OSS project administration.}
% from the theory section
In particular, systems that generate public goods address these and other endemic social challenges by creating governance institutions for attracting, maintaining, incentivizing, and coordinating contributions. 
Ostrom~\cite{ostrom2009understanding} defines institutions as ``... prescriptions that humans use to organize all forms of repetitive and structured interactions ...''. 
Institutions guide interactions between participants in an OSS project, and can be informal such as established norms of behavior, or more formalized as written or codified rules. 
These norms and formalized rules, along with the mechanisms for rule creation, maintenance, monitoring, and enforcement, are the means through which collective action in OSS development occur~\cite{schweik2012internet}, and they can be tiered or nested, as in the context of OSS projects embedded within an overarching OSS nonprofit organization.

Both methods have separately been shown to be utilitarianly describing the state of a process, however, combining the two perspectives has been barely explored. 
In this paper, we undertake a convergent approach, considering from one side OSS projects' socio-technical structure and on the other aspects of their institutional statements.
Our goal is to use these two perspectives synergistically, to identify when they strengthen and complement each other, and to also refine our understanding of OSS sustainability through the two methodological approaches. 
Central to our approaches is the idea that trajectories of individual OSS projects can be understood in the convergent framework through the context provided by similar projects that already are being readily sustained or have been abandoned. 

We leverage a previously published dataset~\cite{yin2021apache} of traces representing OSS developer's day-to-day activities as part of the Apache Software Foundation Incubator (ASFI) project. These developers are a part of projects that have decided to undergo the process of incubation, toward becoming part of the ASF, and benefiting from the services it provides to member projects. 
The dataset includes historical traces and a sustainability label (graduation or retirement) for each project. 
Graduation is an indication of successful incubation and the readiness of a nascent project to join ASF proper, otherwise the project is retired.
\textcolor{red}{In other words and importantly, in this paper, we use the ASFI project outcomes of graduation or retirement as a measure of sustainability of the project. We assume that graduated projects are sustained longer than retired ones, although that might not always be the case}\footnote{For example, it could be that some ASFI retired projects simply could not adapt to the policies and requirements set in the ASFI program but yet continue on, "in the wild" or perhaps aligned with a different OSS foundation. }.
\textcolor{red}{But key hurdles that OSS projects have to demonstrate to graduate is that they can (1) produce new releases, and (2) show the ability to attract new developers. Both of these factors arguably are key toward the sustainability of OSS projects."}

We utilize this dataset to study the extent to which graduated and retired projects differ from each other, from the point of view of both the socio-technical structure and the institutional governance.
On the socio-technical side, we construct the monthly longitudinal social and technical networks for each project, and calculate several measures describing features of the networks. 
On the institutional governance side, we implement a classifier trained on manual annotations of institutional statements in the publicly accessible email communications among ASF participants. 
Then we compare the findings of our socio-technical and institutional metrics for project-level and individual-level activities.
Next, we perform exploratory data analyses, deep-dive case studies, and eventually, we look at how socio-technical measures associate with the prevalence of institutional statements, and evolutionary trajectories during OSS project incubation to sustainability.
In summary, we find that:

\begin{tcolorbox}[colframe=browna]
\begin{itemize}

\item We effectively extract governance content from email exchange in the form of institutional statements, and they fall into 12 distinguishable topics. 

\item Projects with different \textcolor{red}{graduation (i.e.,} sustainability) outcomes differ in how much governance discussion occurs within their communities, and also in their socio-technical structure. 

\item Self-sustained projects (i.e., graduated) have a more socially active community, achieving it within their first 3 months of incubation, and they demonstrate more active contributions to documentation and more active communication of policy guidance via institutional statements.

\item A project's socio-technical structure is temporally associated with the institutional communications that occur, depending on the role of the agent \textcolor{red}{(mentor, committer, contributor)} communicating institutional statements. 
%\sout{Graduated projects show more unidirectional flow from institutional changes to socio-technical changes.}

\end{itemize}
\end{tcolorbox}

\textcolor{red}{Our study is the first attempt to provide a common framework for simultaneous, socio-technical structure and institutional, analysis of OSS projects, in order to describe and understand a process affected by both, that is, project gaining self-sustaining and self-governing community and eventually graduating from the ASF incubator. 
We are hopeful that refining this convergent approach, of structural and institutional analyses, will open new ways to consider and study emergent properties like project sustainability.}

%\end{document}

\section{Theoretical Framework}
%\documentclass[main.tex]{subfiles}
%\begin{document}

%While organizational and socio-technical perspectives have long dominated the analysis of OSS projects, and the socio-technical side of software engineering generally, they miss some basic tensions that can determine project sustainability.
%\correct{CS: consider replacing "success" with "sustainability" if we are doing that everywhere} or failure 
%\correct{CS: consider replacing "failure" with "abandonment. I think we need to be careful in GCR project papers to always explain what we mean by success (e.g. graduated vs retired; sustained versus abandoned. And we have the complication that retired projects can leave and still be sustained}. 

Here we introduce the theories behind the two different viewpoints, Institutional Analysis and Development (IAD) and Social-Technical Systems (STS), as well as Contingency Theory serving as the glue between institutional governance and the organizational structure of OSS projects.

\subsection{Institutional Theory and Commons Governance}
OSS projects are a form of digital commons, or more precisely, commons-based peer production~\cite{schweik2012internet}. 
Legal scholar Yochai Benkler~\cite{benkler2008wealth} introduced the phrase Commons-Based Peer-Production (CBPP) to describe situations where people work collectively over the Internet, and where organizational structure is less hierarchical. 
Individual participants select for themselves tasks to work on. While CBPP situations are found in a variety of settings, Benkler argues that OSS is the `quintessential instance' of CBPP. 

There is a relatively long history of the study of governance in commons settings, arguably led by Nobel laureate Elinor Ostrom and her groundbreaking book Governing the Commons~\cite{ostrom1990governing}.  
Ostrom's Institutional Analysis and Development (IAD) framework was developed to study the governance institutions that communities develop to self-manage natural resources. 
Much of this research focuses on the governance and sustainability of natural resource settings, e.g., water~\cite{blomquist1992dividing}, marine~\cite{gruby2013multi,lindkvist2017micro}, and forest~\cite{fleischman2014evaluating} settings. 

\textcolor{red}{A key challenge in natural resource commons settings is that individuals who cannot easily be excluded from extracting resources from the pool of available natural resources often have little incentive to contribute toward the production or maintenance of that resource – what are commonly referred to as `free-riders'~\cite{olson2012logic,sandler2004global}. In forest, fishery or water settings, the free-rider problem in open access settings can lead to a problem termed by Hardin as the `Tragedy of the Commons'~\cite{hardin1968tragedy}. 
Ostrom famously pushed back against Hardin’s analysis and over a course of a lifetime of work, highlighted that communities can avoid tragedy through hard work in developing self-governing institutions.}

\textcolor{red}{OSS commons are fundamentally different from natural resources in that digital resources can be readily replicated and are not subject to degradation due to over-harvesting. 
Therefore, if over-appropriation is not a problem, is there a potential tragedy of the commons in an OSS context? Invariably the answer is yes, and it lies at the heart of the idea of OSS sustainability. The tragedy occurs when there are free-riders and insufficient human resources available to continue to further develop and maintain the software and, as a result, the software project fails to achieve the functionality and use that was perhaps envisioned when it began, and becomes abandoned~\cite{schweik2007tragedy}. Ostrom and Hess~\cite{hess2007understanding} aptly describe this tragedy as `collective inaction.'}

\textcolor{red}{Ostrom’s Nobel Prize winning body of work was studying how humans collectively act and craft self-governing institutional arrangements to effectively avoid the tragedy in natural resource settings. 
Central in this effort was the introduction and evolution of the Institutional Analysis and Development (IAD) framework~\cite{ostrom2009understanding}. 
Later, IAD was applied to the study of digital or knowledge commons \cite{GKC2014,hess2007understanding} and explicitly to the study of self-governance in OSS, where Schweik et. al. undertook the first study of technical, community, and institutional designs of a large number of OSS projects~\cite{schweik2012internet}.}

With that being said, prior work has found that self-governing OSS projects develop highly organized social and technical structures~\cite{bird2008}. 
Those having foundation support, like the ASF, may additionally be in the process of organizing the developers' structured interactions under a new set of governance prescriptions as required by the ASF Incubator.
We refer to an individual institutional prescription as an \textit{Institutional Statement} (IS), which can include rules and norms, and which we define as a shared linguistic constraint or opportunity that prescribes, permits, or advises actions or outcomes for actors (both individual and corporate)~\cite{CrawfordOstrom1995,siddiki2019institutional}. 
Institutions, understood operationally as collections of institutional statements, create situations for structured interaction for collective action. In other words, configurations of ISs affect the way collective action is organized. In the context of ASF and OSS projects, incubator ISs can affect OSS project social and technical structure. 

With IS and other approaches to %\sout{institutions}
\textcolor{red}{institutional analysis}, it becomes possible to articulate the relationships between governance, organizational, and technical variables. For example, previous studies on OSS often report code modularity as a key technical design attribute~\cite{o1999lessons, narduzzo2005role}. Hissam et al.~\cite{hissam2001perspectives} write: `A well-modularized system ... allows contributors to carve off chunks on which they can work.'
%\sout{Verbal institutional commitments in ASF projects to openness and transparency, captured in the form of institutional statements, could then}
\textcolor{red}{Open and transparent verbal discussion between OSS team members and other ASF officials (e.g., mentors) about OSS project or ASF institutional design, captured in the form of institutional statements, could then} predict effort by project contributors to restructure their project’s technical infrastructure to be more modular and inviting to new contributors. 
%To this point, commons-based approaches have been applied to the Internet backbone~\cite{Hess1995virtualCPR}, `knowledge commons' such as genomic registries~\cite{GKC2014}, online discussion communities~\cite{kollock1996managing,kraut2012building}, and game communities~\cite{frey2019place,zhong2020institutional},
Using the approaches of institutional analysis, we extract institutional content from open access email exchanges between OSS project contributors to understand the role of communication governance information in OSS project sustainability.

%\textcolor{red}{CS TO LIKANG: NONE OF THE REFERENCES ARE WORKING IN THIS SECTION AND DON'T UNDERSTAND THE PROBLEM. BELOW ARE THE REFERENCES I HAVE HARDCODED IN THE NEW TEXT I HAVE ADDED. CAN YOU ADD THEM OR EXPLAIN TO ME HOW TO FIX THEM IN OVERLEAF? I SEE THE REF.BOB... THANKS.} 

%\textcolor{red}{Brett M. Frischmann, Michael J. Madison and Katherine J. Strandburg (eds). 2014. Governing Knowledge Commons. Oxford, UK: Oxford University Press.} 

%\textcolor{red}{Hardin, Garrett. 1968. “The Tragedy of the Commons.” Science. 162: 1243-1248.}  

%\textcolor{red}{Olson, Mancur. 1965. The Logic of Collective Action. Cambridge, Mass.: Harvard University Press.}

%\textcolor{red}{Ostrom, E , 1990. Governing the Commons: The Evolution of Institutions for Collective Action. Cambridge: Cambridge University Press.}

%\textcolor{red}{Ostrom, Elinor. 2005. Understanding Institutional Diversity. Princeton, NJ: Princeton University Press.} 

%\textcolor{red}{Hess, Charlotte and Ostrom, Elinor (eds). 2011. Understanding Knowledge as Commons. Cambridge, MA: MIT Press.}  

%\textcolor{red}{Sandler, Todd. 2004. Global Collective Action. Cambridge: Cambridge University Press.}

%\textcolor{red}{Schweik, Charles and English, Robert. 2007. “Tragedy of the FOSS Commons? Investigating the Institutional Designs of Free/Libre and Open Source Projects. First Monday. 12 (2). https://firstmonday.org/article/view/1619/1534}

\subsection{Socio-Technical System Theory}
A Socio-Technical System (STS) comprises two entities~\cite{trist1981evolution}: the social system where members continuously create and share knowledge via various types of individual interactions, and the technical system where the members utilize the technical hardware to accomplish certain collective tasks.
The STS theory can be considered to combine the views from both engineers and social scientists, an intermediary entity of sorts, that transfers the institutional influence to individuals~\cite{ropohl1999philosophy}.
The theory of STS is often referenced when studying how a technical system is able to provide efficient and reliable individual interactions~\cite{herrmann2004modelling}, and how the social subsystem becomes contingent in the interactions and further affects the performance of the technical subsystem~\cite{fischer2011socio}.
Moreover, the socio-technical system theory plays an important role in analyzing collective behavior in OSS projects~\cite{ducheneaut2005socialization,bird2006mining,wu2007investigating}.
OSS projects have also been studied from a network point of view~\cite{ducheneaut2005socialization}. Gonz{\'a}lez-Barahona et al.~\cite{gonzalez2004community} proposed using technical networks, where nodes are the modules in the CVS repository and edges indicate two modules share common committers, to study the organization of ASF projects. 
In socio-technical systems, organizations can intervene through long-term or short-term means. Smith et al.~\cite{smith2007moving} propose two conceptual approaches, `outside' and `inside': `outside' approaches represent the socio-technical and are managerial in approach. `inside' approaches are more reflexive about the role of management in co-constituting the socio-technical. 

From that perspective, the Apache Software Foundation (ASF)  community is a unique system that has both outside influence regulations from ASF board and members 
and inside governance managed or self-governed by individual Project Management Committees (PMC).

\subsection{Contingency Theory}
Contingency theory is the notion that there is no one best way to govern an organization. 
Instead, each decision in an organization must depend on its internal structure, contingent upon the external context (e.g., stakeholder~\cite{turner2004communication}, risk~\cite{cooke2002real}, schedule \cite{wearne1989study}, etc.). Joslin et al.~\cite{joslin2016impact} find that project success is associated 
with the methodologies (e.g., processes, tools, methods, etc.) adopted by the project. Here, in particular, we treat the institutional statements as an abstraction of the methodologies in OSS development.
As the organizational context changes over time, to maintain consistency, the project must adapt to its context accordingly. 
Otherwise, conflicts and inefficiency occur~\cite{barclay1991interdepartmental}, i.e., not a single organizational structure is equally effective in all cases. 
Similar arguments have been made in the field of institutional analysis, arguing that there are no panaceas or standard blueprints for guiding the institutional design of a collective action problem ~\cite{OstromJanssenAnderies2007}.

To address the conflicts caused by incompatibilities with the project’s context, previous work suggests thinking holistically. 
Lehtonen et al.~\cite{lehtonen2006three} consider the project environment as all measurable spatio-temporal factors when a project is initiated, processed, adjusted, and finally terminated. They suggest that the same factor can have an opposite influence on the projects under a different context. 
Joslin et al.~\cite{joslin2016impact} consider project governance to be part of the project context, concluding that project governance can impact the use and effectiveness of project methodologies. 
%M{\"u}ller et. al.~\cite{joslin2016impact} further summarize the selection method as a qualitative measure on the relationship between context and performance, while the interaction method works more quantitatively. And the system method addresses the contextual factors (including both temporal and structural variables) holistically.

As per contingency theory, during ASFI projects' incubation, developers and mentors have to make in-time decisions on their organizational structure, contingent on what is happening on the institutional rules and governance, and vice versa.

%\end{document}

\section{Research Questions}
Reflecting on the previous discussion, the primary goal of this paper is to demonstrate that the evolution of a project from a nascent state to a sustainable state can be studied effectively by combining the two different methodologies of socio-technical network analysis and institutional analysis.

We reported in prior sections that a variety of scholars have utilized a socio-technical systems approach to analyze collective behavior in OSS projects. 
We also described how institutional analysis is useful in understanding collective action in OSS settings. 
To enable to dual-view on sustainability, we first describe and evaluate our automated approach to identifying Institutional Statements in project emails.

\textbf{RQ$_{\textcolor{myred}{1}}$: Are there Institutional Statements contained in ASF Incubator project email discussions? Can we effectively identify them?} 

With the next two research questions, we assess the utility of our convergent approach to the institutional analysis and socio-technical system frameworks. 
In the case of the ASF incubation program, there are two eventual outcomes: either a project graduates from the ASF incubator and becomes a full-fledged ASF associated project, or it `retires' without achieving that goal. 
In this context, we operationalize a sustainable state as one where an OSS project graduates from the ASF incubator program, rather than retires.
We ask:

\textbf{RQ$_{\textcolor{myred}{2}}$: Is OSS project evolution toward sustainability readily observable through the dual lenses of institutional and socio-technical analysis? And how do such temporal patterns differ?}

Per institutional analysis theory, strategies, norms, and rules can affect the social and technical organizations of projects.
Governance and organization, per social theories, must work hand-in-hand to make viable socio-technical systems.  Ill-designed institutional arrangements would introduce inefficiencies into the system, and such inefficiencies may amplify deviant behaviors and irregular structure in the systems. 
Such influential links from institutional design to the organizational structure can be, in fact, bi-directional.
In effect, in a sustainable system, an ill-formed organizational structure may instigate new rules to adjust and improve such structure, further improving efficiencies in the systems.

Thus, we hypothesize that the feedback, if any, between project governance and project organization should be observable, specifically in that intensified governance discussion should precede and/or follow changes to the project organizational structure.
As a reminder, we consider Institutional Statements as indicators of intensified discussions of OSS project self-governance or new incubator requirements on that self-governance. We also consider socio-technical network parameters as indicators of organizational structure.
Thus, we ask:

\textbf{RQ$_{\textcolor{myred}{3}}$: Are periods of increased Institutional Statements frequency followed by changes in the project organizational structure, and vice-versa?} 

In the following section, we introduce the methodologies approaching the above three research questions.

\section{Data and Methods}
%\documentclass[main.tex]{subfiles}
%\begin{document}
\textcolor{red}{To study the difference between projects that graduate ASFI (i.e., become sustainable) and those that do not, in this paper we use a collection of large-scale data sets comprising Institutional Statements and Socio-Technical variables extracted from all graduated and retired projects from the Apache Software Foundation Incubator, ASFI. In ASFI, graduation is an indication that a nascent project is sufficiently sustainable
to join ASF proper\footnote{ASF's guide to project graduation: \url{https://incubator.apache.org/guides/graduation.html}}, otherwise the project is retired.
Our combing through the Apache lists, inspecting the data, and speaking to project and community members have shown that almost all failures to graduate are sustainability failures. 
In rare occasions, some projects have retired for reasons other than sustainability, e.g., some are not a good fit for the Apache model\footnote{ASF's reason behind projects' retirement: \url{https://incubator.apache.org/projects/\#retired}},  despite evidence that projects are generally sufficiently aware of the ASF model before entering incubation according to their project proposal\footnote{ASF incubator projects' proposal \url{https://cwiki.apache.org/confluence/display/INCUBATOR/Proposals}}.}

%\textcolor{red}{For large-scale data efforts like ours, integrating IAD and STS approaches require heterogenous data sets and automated approaches.}

For the socio-technical networks, we collected historical trace data of commits, emails, and incubation outcomes for 253 ASFI projects, which have available archives of both commits and emails from 03/29/2003 to 02/01/2021\footnote{Our code and data is available at Zenodo: \url{https://doi.org/10.5281/zenodo.5908030}}. 
Among those, 204 projects have already graduated, and 49 have retired.
ASF incubator projects that are still in incubation are not studied in this paper.

We collected the ASF incubator project data from the ASF mailing list archives\footnote{During the submission of this study, ASF has moved their email archives to Pony Mail.}, which are open access and can be retrieved through the archive web pagelists, \url{http://mail-archives.apache.org/mod_mbox/}.
They contain all emails and commits from the project's ASF incubator entry date, and are current. 
The project URLs follow the pattern: \texttt{proj\_name - list\_name}\texttt{/(YYYYMM).mbox}.
For example, the full URL for the {\it dev} mailing list of the \emph{Apache Accumulo} project, in Dec 2014, is \url{http://mail-archives.apache.org/mod\_mbox/accumulo-dev/201412.mbox}. 
Each such $.mbox$ file contains a month of mailing list messages from the project, for the date specified in the URL. 
Here {\it dev} stands for `emails among developers'. 
Notably, there are some sites that are not following the pattern, e.g., 'ASF-wide lists' are not project-owned mailing lists, and list 'incubator.apache.org' contains data of more than one project.

\textcolor{red}{To extract Institutional Statements, we combined our email data set with a prior data set on ASF policy documents.
In a given organization, institutional statements are characterized by a finite set of semantic roles (e.g. ASF Board, Mentors, contributors, etc. \textcolor{red}{in ASF}), and their interactions (e.g. management committees requesting reports from projects, developers voting to induct committers \textcolor{red}{in ASF}), in specific contexts.}
\textcolor{red}{To account for their representation in our training corpus, we included institutional statements from not only ASF project-level email exchanges among participants, but also ASF policy documents.}
\textcolor{red}{The supplementary set of Institutional Statements included 328 policies, which were compiled from ASF policy documents (e.g., Apache Cookbook, PPMC Guide, Incubator Policy, etc), in an economic analysis of the ASF Incubator's policies~\cite{sen2021cui}.}

%We also collected all other mailing lists that currently exist, e.g., `commits' (committers commit information), `issues', `notifications', `users' (emails between users and developers or other users).
%Moreover, we find that many ASFI projects, especially these over ten years old, which used SVN, used a bot in the `dev' mailing list to record all commits, thus a message from `dev' is not always an email from real person. 
%Similar emails might be sent to the `commits' mailing list, which, thus, contains some emails. 
%Therefore, we parse both `dev' and `commits' mailing lists files for commits information and emails information in the 233 ASF incubator projects through the archive web page\footnote{Data and scripts are available: X}. In total, we identify 114,584 .mbox files across all existing mailing lists from all incubator projects.

\subsection{Pre-processing}
\textcolor{red}{We collected all 1,330,003 emails across the ASF Incubator projects, from 03/29/2003 to 02/01/2021 (under mailing lists of `dev', `user', `announce', etc.). 
We find that 128,257 (about 9.6\%)} emails are automatically generated and broadcast by continuous integration tools \textcolor{red}{(i.e., bots)}. 
Because the amount of such emails is substantial, \textcolor{red}{but} they carry less meaningful social or institutional information, and list members almost never reply to them, we use regular expression rules to identify and eliminate them from the corpus, \textcolor{red}{leaving us 1,201,746 emails.}

And, for the technical contribution side, many projects, especially those over ten years old that used SVN, utilized a bot for extensive mailings, thus forming outliers in the dataset. 
Thus, we eliminate commit messages from automated bots (e.g., `buildbot'), 253,758 out of 3,654,196 (about 14.4\%) commit messages, and email messages from issues/bug tracking bots (e.g., `GitBox').
Moreover, we find some developers contributed commits by directly changing/uploading massive non-source code files (e.g., data, configuration, and image files).
Since committing non-coding files can form outliers in the data set, we choose to apply the {\it GitHub Linguist}\footnote{GitHub Linguist~\url{https://github.com/github/linguist}} to identify 731 collective programming language and markup file extensions, and exclude any other non-coding commits (e.g., creating/deleting folder, upload images, etc.). 

\subsection{Constructing Socio-technical Networks}  
Network science approaches have been prominent in studying complex systems, e.g., OSS projects~\cite{bird2009putting,surian2013predicting}.
Since networks can contain rich information for both the elements (i.e., nodes) and their interactions (i.e., edges), in this study, we use socio-technical networks to anchor the abstraction of socio-technical systems.
We define the projects' socio-technical structure using social (email-based) and technical (code-based) networks, extracted from their emails to the mailing lists and commits to source files.
Similar to the approach by Bird et al.~\cite{bird2006mining}, we form a social network (weighted directed graph) for each project, at each month in their incubation, from the communications between developers: a directed edge from developer \textit{A} to \textit{B} forms if \textit{B} has replied to \textit{A}'s post in a thread or if \textit{A} has emailed \textit{B} directly. The weight of the edge represents the communication frequency between a pair of developers.
The technical bipartite networks (weighted bipartite graph) are formed in a similar way. For each project, at each month, we include an un-directed edge between a developer \textit{A} and a source file \textit{F} if  developer \textit{A} has committed to the source file \textit{F} that month (excluding the SVN branch names). The weight of the edge represents the committing frequency between the developer and the source file.
In summary, social networks are weighted directed graphs. We form edges between two developers nodes, if one developer replied to or referenced the other's email.
Technical networks are undirected bipartite graphs, with developers forming one set of nodes, coding files forming the other, and a link being drawn when a developer contributed to a coding file. 
We use the \textit{Python} \textit{networkx} package for the network-related implementation.

\subsection{Extracting Institutional Statements}
 
 \textcolor{red}{We combined the email exchange data set with the ASF policy document data to fine-tune a BERT-based ~\cite{cohan2019pretrained} classifier, for automatic detection of ISs (see Sect. 2.1 for definition of IS).} 

\textcolor{red}{To start, we hand-annotated a small subset of our data for ISs as follows\footnote{The coding manual for training the annotators can be found here: https://doi.org/10.5281/zenodo.5908030.}. %The  corpus of institutional statements from ASF project emails was drawn from a hand-labeled dataset that we built as follows. 
After selecting a random subset of 313 email threads from incubator project lists, two hand-coders labeled the sentences in them as "IS" or "Not IS", on the basis of whether they fit the definition of Institutional Statements.
They resolved disagreements through discussion and recorded these conclusions, achieving a peak out-of-sample agreement between $0.75$ to $0.80$.}
A sentence was \textcolor{red}{coded as an IS} only if it was a complete sentence; 
fragments such as parenthetical mentions of rules or resources were not annotated as positive. 
This resulted in \textcolor{red}{6,805} labeled sentences (i.e., "IS" or "Not IS"); \textcolor{red}{273} were labeled as IS.
\textcolor{red}{We present three example ISs in Table~\ref{table:IS}.}

\textcolor{red}{We treated all 328 policies from the ASF documents as institutional statements, since policy documents provide arguably more formal institutional sample text compared to the norm in the email discussions.
%, and therefore likely to enhance the robustness of the model. We treated these \textcolor{red}{328} policies found as institutional statements akin to those expressed via email.
Thus, we had \textcolor{red}{601} Institutional Statements in total across these two coded datasets.}

\begin{table}[b] \centering
  \caption{Selected Examples of Institutional Statements Found in ASFI Project Email Discussions.} 
  \label{table:IS} 
\scalebox{0.85}{
\begin{tabular}{ |  p{0.1\linewidth} |  p{0.15\linewidth}|  p{0.75\linewidth}| }
\hline
Project & Date & Institutional Statements \\ 
\hline
Airflow \linebreak & 21 Dec 2016 & \uline{... running in our Lab there is virtually no restriction what we could  do,  however I will hand select people who have access to this environment. I will also hold ultimate power to remove access from anyone ...} \\ 
\hline
%NPanday \linebreak (users mailing list) & 02 Feb 2011 & 
%... \uline{You can check out the docs here: $<URL>$} \\
%\hline
ODF \linebreak & 07 Dec 2011 &  \uline{Please vote on releasing this package as $<Package>$. The vote is open for the next 72 hours and passes if a majority of at least three +1 ODF Toolkit PMC votes are cast ...} \\ 
\hline
 Airflow\linebreak  &  24 Feb 2017 & \uline{... Next steps: 1) will start the voting process at the IPMC mailinglist}. ... \uline{So, we might end up with changes to stable.} ... \uline{2) Only after the positive voting on the IPMC and finalisation I  will rebrand the RC to Release.} \\
\hline
\end{tabular}} 
\end{table}

%\textcolor{blue}{Below we switch to threads but above we said 1000 random emails. Which one is it? We have to be consistent.}

% 

%Segments were labeled 
%\textcolor{red}{The annotation of sentences into IS or Not IS was performed by two authors, who coded independently and resolved differences jointly, with our discussions carefully documented.}

%contained ISs. We did not include more than one segment per email, and we only included a segment if it% 

%\textcolor{red}{In additiona to the email threads, we also used ASF policy documentations [WHICH ONES? HOW MUCH?], which are an expansive record of precedents and recommendations used in project administration. 
%The institutional statements in those were carefully extracted by domain experts from ASF manuals, bylaws and guides (unpublished). They are included as training data augmentation to familiarize classifiers with semantic roles and information which characterize institutional statements (namely agents, actions and context).}

\underline{BERT-based Sequential Classifier} In natural speech, such as emails, ISs can appear as whole sentences, parts of sentences, or span multiple sentences. 
They are also relatively sparse, with their institutional quality dependent on their inherent interpretation as well as context. 
Framing IS extraction as a sequential sentence classification task in the context of self-contained email segments, instead of labeling individual sentences helps take into account contextual cues. 
\par \textcolor{red}{We used the sequential sentence classifier developed by Cohan et al.~\cite{cohan2019pretrained}, which leverages Bidirectional Encoder Representations from Transformers (BERT) sequence classifier~\cite{devlin2018bert} to classify sentences in documents. As demonstrated there BERT can be employed to generate representation for a sentence in a document's context, through joint encoding over it's neighbouring sentences and then leveraging the corresponding sentence separator '<SEP>' token's tuned embedding for downstream applications, such as sentence labeling, extractive summarizing, etc. Thus, our classifier comprises of BERT for attention-based joint encoding across sentences followed by a feedforward classifier to predict sentence labels based on these separator '<SEP>' vectors. }
\par \textcolor{red}{To test the performance of the classifier on email IS extraction, we held-out 40 email threads (12.5\%, randomly split) out of our 313 hand annotated email threads. The training was performed on the combined set of the remaining 273 coded email threads and the ASF policy documents. The coded training and, respectively, testing email data contained 231 and, respectively, 42 institutional statements. For both training and testing, email threads were processed to generate classifier inputs as follows. To include neighboring context while meeting length limits of the BERT-based text classifier, for each email document, sentences were first chunked into segments using a sliding window of up to 256 BERT sub-word (wordpiece) tokens.
This resulted in segments containing 6 contiguous sentences each, on average, comprising as many full sentences as could be accommodated in the specified subword limit. The rolling window had a step of 1 full sentence. We generated 3322 and 384 email segments for training and testing, respectively. For the policy documents, each policy with its sentences was treated as a segment, leading to 328 additional segments in the training data.}
\par \textcolor{red}{We fine-tuned our classifier end-to-end against the corresponding labels for sentences in the segment. Training was conducted with a batch size of 16 and learning rate of $2\cdot 10^{-5}$, for 6 epochs.
%on a single 16 GB GPU. 
All other hyperparameters were left as defaults. To account for the class imbalance, we randomly oversampled training data segments which had at least one IS sentence to match the number of segments which had no IS sentences (1:1). In both the training and predicting phase, we did not incorporate any temporal information, other than the sequentiality captured by the segments. That is, when extracting the institutional statements, the model does not require the exact time of the discussion. }
\par During testing or prediction, due to variable length of context preceding or following each sentence in any particular segment, we treat a sentence in an email as a "positive" classification, if it has been detected as an IS in at least one segment. The performance of the model has been reported in terms of the F1-score, precision and recall with respect to the positive (`IS') label detected for sentences in the test email set in Sect. 5.1.

\subsection{Topics Identification in Institutional Statements}
The purpose of text modeling is to describe the text given a specific corpus, and provide numerically measurable relationships among texts, e.g., topics identification, measuring similarity, etc.
We use a Latent Dirichlet Allocation (LDA) model to get semantically meaningful topics to better understand the extracted institutional statements. 
\textcolor{red}{LDA is an unsupervised clustering approach~\cite{yu2001direct}, which when given a set of documents, iteratively discovers relevant topics present in them, based on the  word distributions and relative prevalence in each document.
We used LDA to identify prominent topic clusters occurring among all institutional statements extracted from our email archives through our trained classifier (see Sec 4.3).
No prior training from our coded email set against pre-identified topic labels was used to train the LDA model.}
To optimize the number of topics in the LDA model, we use topic coherence score provided by Python $Gensim$ package~\cite{vrehuuvrek2011gensim}.

\subsection{Variables of Interest}
\label{variables}
We draw institutional and socio-technical project features and variables on the basis of each framework's predictions for our research questions. 
\textcolor{red}{Our socio-technical variables are pulled from a recent study on forecasting the sustainability of OSS projects~\cite{yin2021sustainability}, showing high predictive power of socio-technical variables.}
All metrics are aggregated over monthly intervals, for each project, from the start to the end of its incubation.

\begin{table}[h] \centering 
  \caption{Summary statistics for the \textcolor{red}{monthly} socio-technical variables and \textcolor{red}{the counts of institutional statements from project mentors, committers, and contributors} after removal of the top 2\% of outliers. 
  \textcolor{red}{The numbers in parentheses denote the values after the removal of inactive months (i.e., absent of emails/commits).}
  Prefix \texttt{s\_} denotes features in the social network while \texttt{t\_} represents the technical network.} 
\scalebox{.95}{
\begin{tabular}{@{\extracolsep{5pt}}lHccHccH} 
\\[-1.8ex]\hline 
\hline \\[-1.8ex] 
Statistic &  & \multicolumn{1}{c}{Mean} & \multicolumn{1}{c}{St. Dev.} &  & \multicolumn{1}{c}{25\%} & \multicolumn{1}{c}{75\%} & \\ 
\hline \\[-1.8ex] 

s\_num\_nodes & 3 218 & 13.04 (16.96) & 14.56 (15.04) & 2 & 4 (7) & 17 (22) & 98 \\ 
s\_graph\_density & 3 218 & 0.30 (0.30) & 0.27 (0.22) & 0.03 & 0.12 (0.14) & 0.40 (0.40) & 1.00 \\ 
s\_avg\_clustering\_coef & 3 218 & 0.22 (0.29) & 0.23 (0.21) & 0.00 & 0 (0.11) & 0.39 (0.43) & 1.00 \\ 
s\_weighted\_mean\_degree & 3 218 & 11.83 (15.56) & 12.03 (12.81) & 2.00 & 4 (7.43) & 16 (19.71) & 137.45 \\ 
\hline
t\_graph\_density & 3 218 & 0.37 (0.68) & 0.41 (0.32) & 0.11 & 0 (0.36) & 1 (1) & 1 \\ 
t\_num\_dev\_nodes & 3 218 & 1.18 (2.21) & 1.59 (1.60) & 1 & 0 (1) & 2 (3) & 9 \\
t\_num\_file\_nodes & 3 218 & 60.99 (114.83) & 153.94 (197.25) & 1 & 0 (6) & 38 (126) & 1 229 \\ 
t\_num\_file\_per\_dev & 3 218 & 28.79 (53.57) & 80.46 (104.23) & 1 & 0 (4) & 20 (54.5) & 1 174 \\ 
\hline
num\_IS\_mentor & 3 218 & 15.46 (15.99) & 24.46 (25.01) & 0 & 0 (1) & 20 (20) & 294 \\ 
num\_IS\_committer & 3 218 & 9.34 (12.89) & 19.36 (22.36) & 0 & 0 (0) & 10 (16) & 455 \\ 
num\_IS\_contributor & 3 218 & 13.18 (16.36) & 21.72 (24.42) & 0 & 0 (2) & 18 (21) & 432 \\ 
\hline \\[-1.8ex] 
\end{tabular}}
\label{stats}
\end{table} 

\uline{Longitudinal Socio-Technical Metrics}:  
For each project network, for each month, we constructed the social and technical networks, and from them calculate various organizational structure measures. 
In our tables and results, the prefix \texttt{t\_} in a variable's name indicates it is of the technical (code) network, while the prefix \texttt{s\_} in a variable's name indicates it is of the social (email) network. 
For the monthly social networks, we calculate the weighted mean degree \texttt{s\_weighted\_mean\_degree} (sum of all nodes' weighted degree divided by the number of nodes), average clustering coefficient \texttt{s\_avg\_clustering\_coef} (the average ratio of closed triangles over open triangles), graph density \texttt{s\_graph\_density}.
%, and the size of largest connected component \texttt{s\_largest\_component}.
In the technical bipartite networks, for each month, we calculate the number of unique developer nodes \texttt{t\_num\_dev\_nodes}, \textcolor{red}{the number of unique file nodes \texttt{t\_num\_file\_nodes}}, the number of files per developer \texttt{t\_num\_file\_per\_dev}, and the graph density \texttt{t\_graph\_density}. 

\uline{Institutional Statements Frequency Metrics}:
For each project, in each month, we added up the ISs in all emails in that month sent by each of the following three separate and identifiable groups of people: 
ASF mentors (\texttt{num\_IS\_mentor}), 
registered ASF committers (\texttt{num\_IS\_committer}), and contributors (\texttt{num\_IS\_contributor}). 
We summarize their statistics in Table~\ref{stats}.
We note that there is a final group of emails, not accounted here, sent by bots. 
Similar to calendar entries, they may be useful, but are not the object of our study here.
%\textcolor{red}{We also test out the proportion of IS in our early experiment setup. However, in the ASF project data, when it comes to some extreme cases, even the same ratio can carry different meanings and further form outliers. This confirms that the absolute numbers carry more accurate information rather than proportions.}

\subsection{Granger Causality}
Time series data allows for the identification of relationships between temporal variables that go beyond association.
One approach, {\em Granger causality}, is a statistical test for identifying quasi-causality between pairs of temporal variables~\cite{dumitrescu2012testing}. 
%\note{CS: citation missing}
Given two such variable, $X_t$ and $Y_t$, the Granger causality test calculates the p-value of $Y_t$ being generated by  a statistical model including only $Y$'s prior values, $Y_{t-1}, Y_{t-2},$ etc., versus it being generated by a model that in addition to $Y$'s prior values, also includes $X$'s prior values $X_{t-1}, X_{t-2}$.
Thus, Granger causality simply compares a base model involving only $Y$ to a more complex model involving $Y$ and $X$, and calculates if the latter is a better fit to the data.
In the context of Granger causality, prior values are called {\em lagged values}, with $X_{t-1}$ having a lag of 1, $X_{t-2}$ having a lag of 2, etc.
If the Granger causality test returns a small enough p-value (e.g., $<0.01$) it is interpreted as rejection of the null hypothesis, and thus establishing that $X$ {\em Granger causes} $Y$.

The Granger causality test makes an assumption that the time-series on which it is applied are stationary, \textcolor{red}{meaning they do not have trend or seasonal effects}.
Therefore, it is necessary to test for stationarity before running the Granger causality.
We use the augmented Dickey-Fuller test~\cite{cheung1995lag}, as implemented in $adf.test$ of the R package $tseries$~\cite{lopez1997power}.
%\note{CS: citation missing}
All of our time-series were found to be stationary.
We note that a distinction is typically made between scientific causality based on controlled experiments, and Granger causality, with the latter only satisfying one (precursor property) of multiple different properties of causality.
Because of that, when Granger causality is used, the word `causality' is always preceeded by `Granger'. 
We also note that this test does not identify the sign, if any (i.e., positive or negative) of the Granger causality. It simply says if one exists.
In this paper we use the $pgrangertest$ function to test Granger causality in $R$. 

%\end{document}

\section{Results}
%\documentclass[main.tex]{subfiles}
%\begin{document}

In this section, we answer the proposed research questions by adopting a dual-view, from the institutional analysis and socio-technical network perspectives.
We first establish the utility of our IS identification methodology.
%We split RQ1 into two questions.

%e would like to first validate the IS technology,

\subsection{\textbf{RQ$_{\textcolor{myred}{1}}$}: Are there institutional statements contained in ASF Incubator project discussions? If any, can we effectively identify the content of ISs?}

\indent \underline{Precision and Recall for Detecting Institutional Statements.} 
First, we focus on the ability of our BERT-based classifier to identify institutional statements in the emails. 
\textcolor{red}{When tested on the 857 held out sentences from the 40 email threads in our test set, see Sect 4.3, } our classifier achieved a \textcolor{red}{precision score of $0.667$, recall score of $0.681$, F1 score of $0.674$, and accuracy of $0.965$ on classifying Institutional Statements, 
%and an overall accuracy of 96.53\% for predicting labels across all sentences.}, 
demonstrating it is able to extract ISs from developer email exchanges in spite of there being only 5.1\% ISs.}
 \textcolor{red}{We consider these performance results satisfactory given that we had a small and highly imbalanced data set (273 ISs out of 6,805 sentences). 
 There are strong indications that increasing the positive examples in the training data set will further increase our classifier's performance.}\footnote{\textcolor{red}{When we fine-tuned the classifier with only the 273 training email threads (i.e., without Institutional statements from the ASF policy documents), the F1 for positive label was found to be about 20\% lower.}}

\textcolor{red}{We ran our classifier on the full corpus of \textcolor{red}{1,201,746 emails (after bot email removal) }} across all ASF incubator projects. 
It identified \textcolor{red}{313,140} ISs in the emails, for an average of \textcolor{red}{0.261} sentence-level ISs per email.
Table~\ref{stats} shows descriptive statistics for both the socio-technical variables and the number of institutional statements from project mentors, committers, and contributors, \textcolor{red}{calculated} in monthly intervals, per project.

We find that the classifier's errors are  also informative. 
In one set of false positives, \textcolor{red}{participants} described plans for an event occurring outside of Apache and the relevant incubator project, not the kind of process or behavioral constraint typical of ISs. 
It was probably detected as an IS due to its semantic similarity to rules and guidelines which make up other positive examples. 
Conversely, the sentence `\textit{Send it to <EMAIL> and see what the reaction is}' was missed as an IS, despite appearing in the context of contributor agreements. 
This miss is likely due to the fact that many such recommendations are made in the emails that would not be considered institutional, because they indicate a particular individual as an individual, rather than in their institutional role.

%\underline{\sout{Power to Describe Content.}} \sout{Here we present our IS content analysis. We consider segments from emails which were identified as containing institutional statements by our classifier, and extracted from them the most frequent non stop-words, in different time intervals with respect to the ASFI exit date. In order of decreasing frequency of appearance, we list the top 10 most frequently used words, in the one month period leading up to and following their exit date, i.e., the graduation or retirement date. Self references to projects themselves were removed from the segments, along with URL and email addresses.}

\begin{figure*}[tbp]
\centering
%\subfigure[Num. IS from Committers ($p < .001$)]{
%\label{num_is_committer} 
%\includegraphics[width=0.23\linewidth]{figs/committer_IS.pdf}}
\subfigure[Num. Mentors IS ($p < .001$)]{
\label{mentor_IS}
\includegraphics[width=0.3\linewidth]{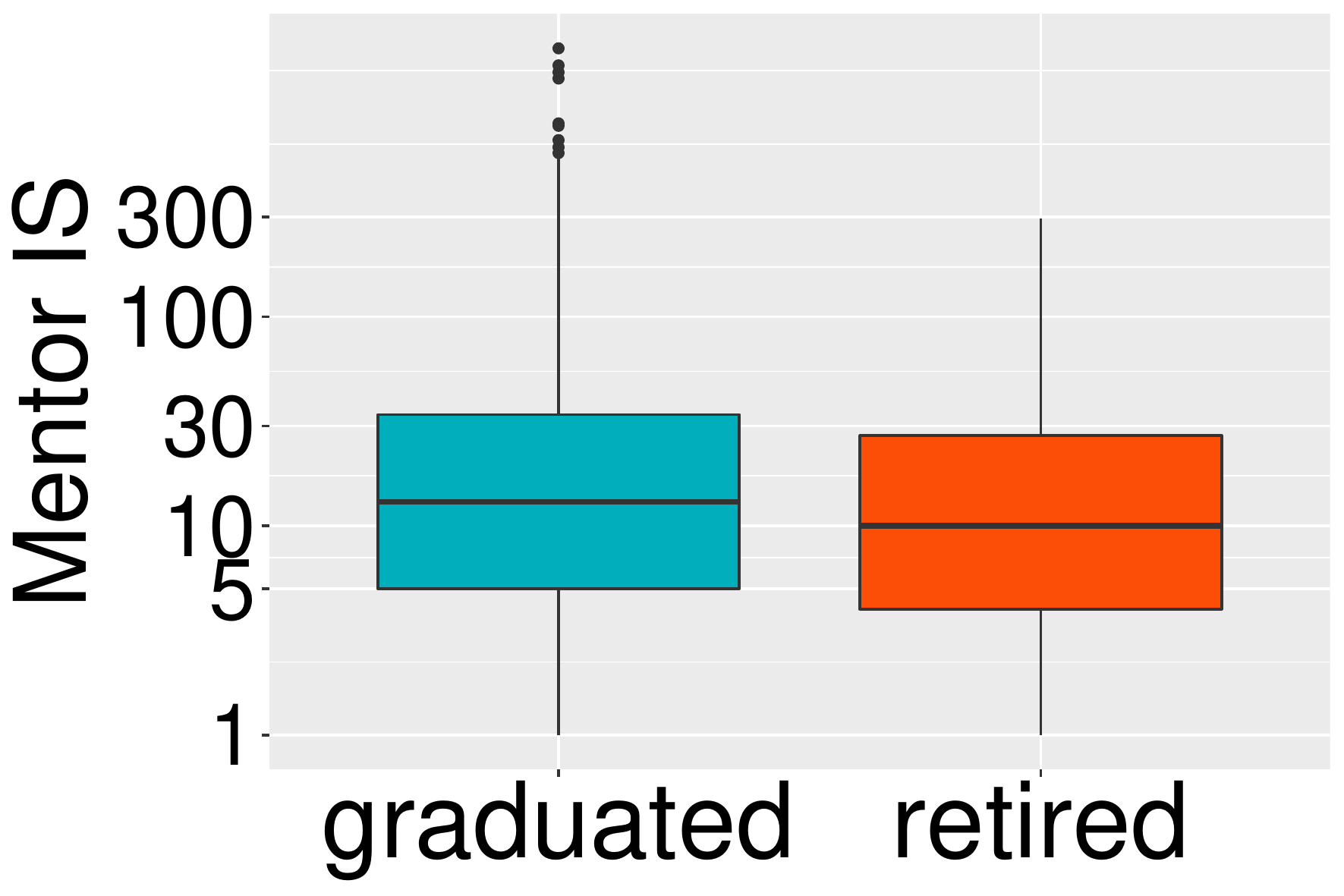}}
%\subfigure[Num. Committer IS per Email ($p < .001$)]{
%\label{num_is_committer_per_email} 
%\includegraphics[width=0.23\linewidth]{figs/committer_per_email.pdf}}
\subfigure[Num. Committers IS ($p < .001$)]{
\label{committer_IS}
\includegraphics[width=0.3\linewidth]{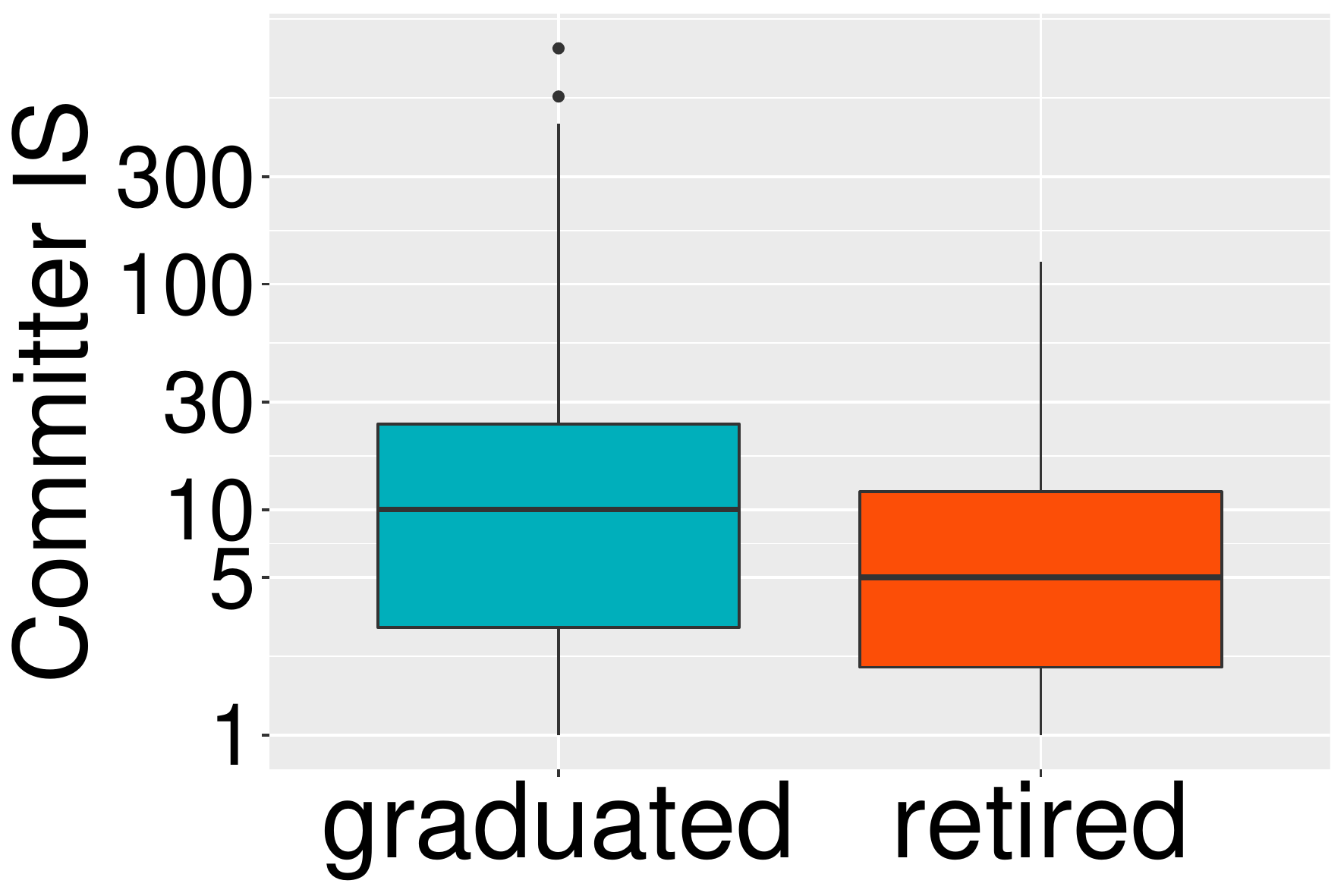}}
\subfigure[Num. Contributors IS ($p < .001$)]{
\label{contributor_IS}
\includegraphics[width=0.3\linewidth]{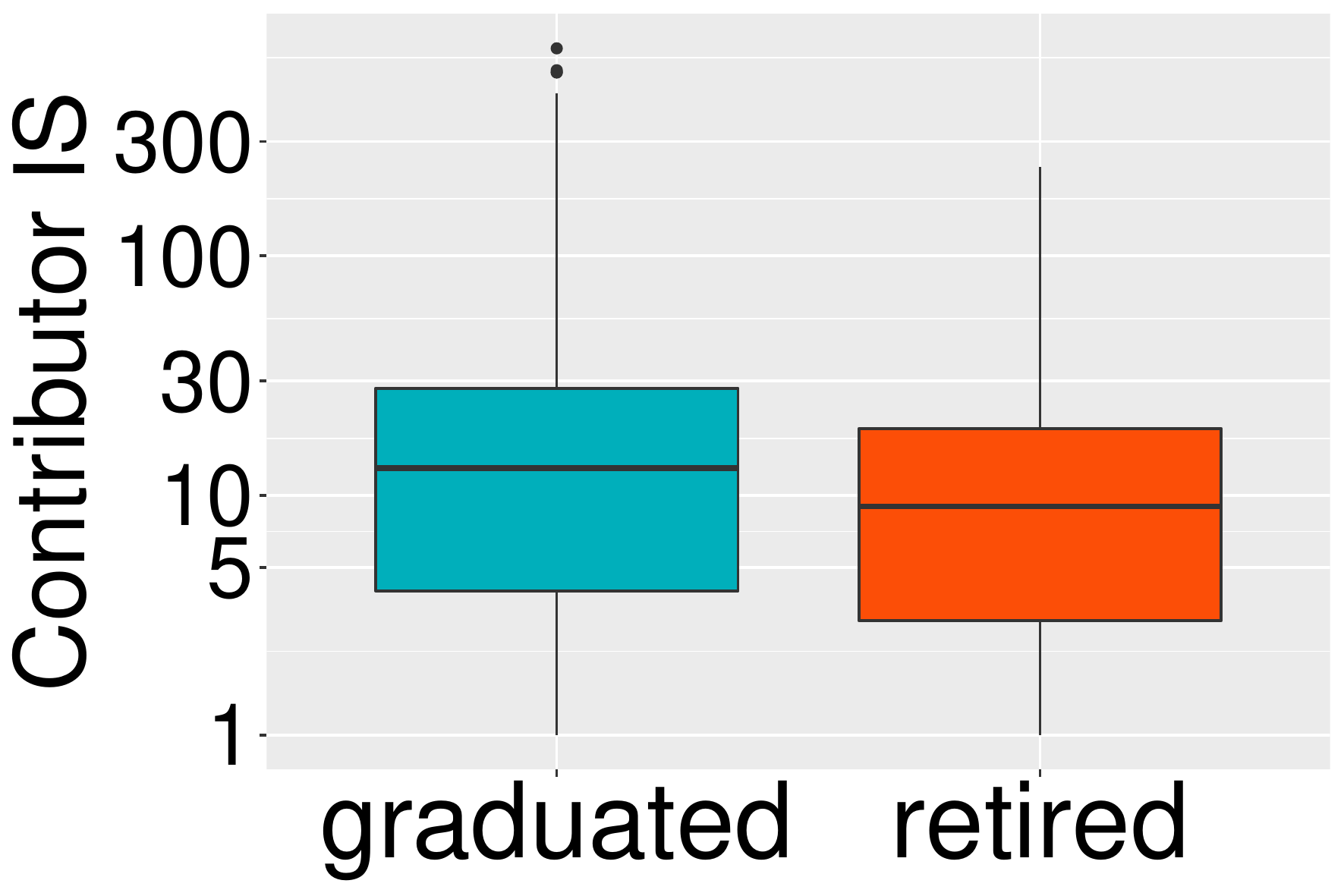}}
%\subfigure[Num. file nodes in technical networks ($p < .001$)]{
%\label{num_file_node_technical_net}
%\includegraphics[width=0.23\linewidth]{figs/t_num_file_nodes.pdf}}
\caption{Comparing graduated vs retired projects along \textcolor{red}{the number of Institutional Statements (IS)}. \textcolor{red}{The Mann-Whitney U test p-values are sufficiently small, suggesting significant differences between groups.}}
\label{fig:boxplots}
\end{figure*}

\underline{Institutional Statements Over Roles and Sustainability Status.}
We turn to some exploratory analysis, to demonstrate the utility of our chosen features when reasoning about differences between graduated and retired projects. 
%\underline{Institutional Statements Metrics} 
Comparing graduated and retired projects, we find a significant difference in the number of ISs. For example, in Figure~\ref{mentor_IS}, the number of IS sent by mentors graduated projects is statistically higher than retired projects \textcolor{red}{(we used the Shapiro-Wilk test to confirm the data is sufficiently normal, and the Mann-Whitney U test for the difference in means)}. This, along with the fact that graduated projects tend to be more active socially overall compared to retired projects (i.e., more email exchanges), suggesting the mentors of retired project are concerned about the projects' community progressing, thus, most of the email content is about rules and guidance.
On the other hand, it is also plausible that mentors engage more socially and less institutionally with graduated projects, which may benefit those projects more.
The numbers of ISs sent by committers and contributors, show similar patterns. 
We investigate them longitudinally in next section.

\begin{table}[tb]
\caption{Topics Identified in Institutional Statements.}
\label{tab:topics}
\begin{tabular}{@{}lll@{}}
\toprule
ID & Heuristic Topic              & Top Sample Words                                        \\ \midrule
1  & Progress Report    & review, require, meeting, board, submit, report         \\
2  & Collective Decision & vote, start, proposal, thread, close, day, bind         \\
3  & Project Release    & release, issue, think, fix, branch, policy              \\
4  & Community & project, email, send, community, behalf, incubation, talk, develop            \\
5  & Report Review      & board, report, time, meeting, prepare, reminder, review \\
6  & Mailing List Issues    & list, mailing, discussion, question, issue, comment, request \\
7  & Documentation      & update, wiki, page, website, documentation, link, doc   \\
8  & Software Testing   & release, source, build, test, note, artifact, check           \\
9  & licensing Policy      & license, file, software, version, copyright, compliance           \\
10 & Routine Work     & project, committer, help, work, way, code               \\
11 & Mentorship         & podling, report, form, mentor, know, sign, month, wish  \\
12 & Software Distribution  & work, repository, information, file, distribute, commit           \\ \bottomrule
\end{tabular}
\end{table}

\underline{Topics Identification in Institutional Statements.}
We use Latent Dirichlet Allocation (LDA) model to study the token-level topics in institutional statements. 
By optimizing the LDA coherence score, we get the optimal number of topics of $12$. 
The result further enables us to study which words are important to each topic.
We present the clusters of top words for each topic in Table~\ref{tab:topics}.

As this table reveals, words are well extracted from the institutional statements and are distinguished from each other. 
For example, in the first topic (i.e., `Progress Report'), there is the cluster of words -- `review', `board' (which relates to ASF board), `submit' and `report' -- all of which are associated with the important incubator rule that requires podlings (i.e., ASF's referring to projects) to report regular progress reports. 
While in topic 7, words like `update', `wiki', `page', `website', and `documentation' emerge, all related to requirements projects need to address related to their website or documentation requirements. 
The results advance the institutional theory under the software engineering domain, arguably that the IS is associated with OSS sustainability, suggesting us to dive deeper into the connections between social-technical system and institutional analysis.

\begin{tcolorbox}[colframe=browna]
\textbf{RQ$_{\textcolor{myred}{1}}$ Summary:} We demonstrated that institutional analysis methodologies can capture differences between graduated projects and retired projects. 
We also showed that we can effectively identify meaningful institutional statements, and common  topics, from ASF incubator projects' emails. 
\end{tcolorbox}

\subsection{\textbf{RQ$_{\textcolor{myred}{2}}$: Is OSS project evolution toward sustainability observable through the dual lenses of institutional and socio-technical analysis? And how do such temporal patterns differ?}}

In this section, our goal is to contrast graduated and retired projects over time in both IS space and socio-technical space. 
Projects exit the ASF incubator at different times. 
In effect, there will be larger variance during the end of the incubation month.
Therefore, we restrict ourselves to the first 24 months for all projects (more than 60\% projects stayed within 24 months in the incubator).

\begin{figure}[tb]
    \centering
    \includegraphics[width=\linewidth]{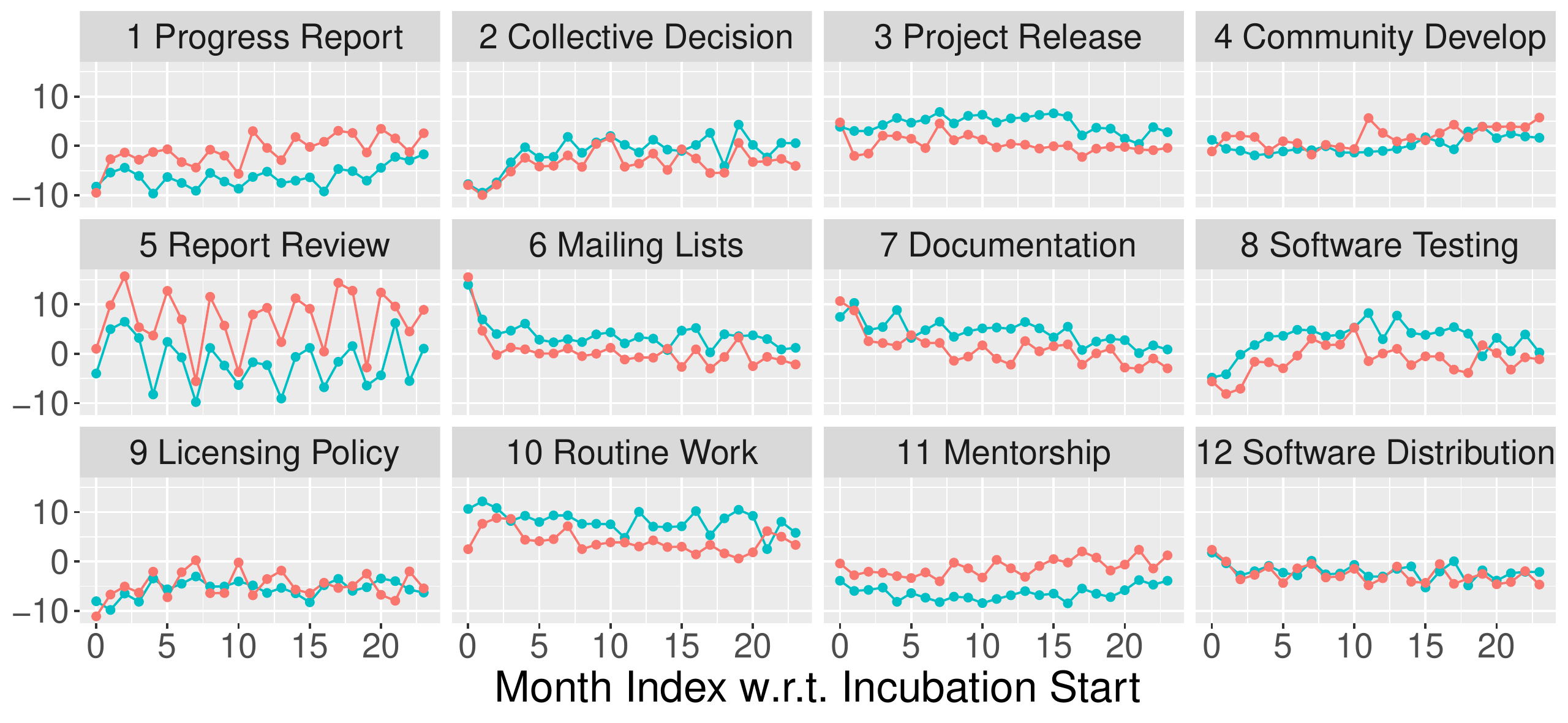}
    \caption{Topics Evolution for graduated projects (in blue) compared to retired projects (in red). X-axis indicates the i-th month from their incubation start, y-axis represents the relative volume of the topics.}
    \label{fig:topicsvs}
\end{figure}

\underline{Topic Evolution Over Time.}
After identifying the words that contribute to various identified topics, by aggregating over all projects, we get the volume, which is measure by the number of tokens contributing to that topic, of each topic in each month. 
Moreover, since there exist trends of the number of IS, we subtract the mean volume for each month, separately for the graduated and retired projects. 
We present them in Figure~\ref{fig:topicsvs}, 
where the x-axis is the number of months after their incubation start, and y-axis indicates the relative volume compared to the mean.

\begin{figure*}[tb]
\centering
\subfigure[Num. IS from Mentors]{
\label{IS_mentors_curve}
\includegraphics[width=0.3\linewidth]{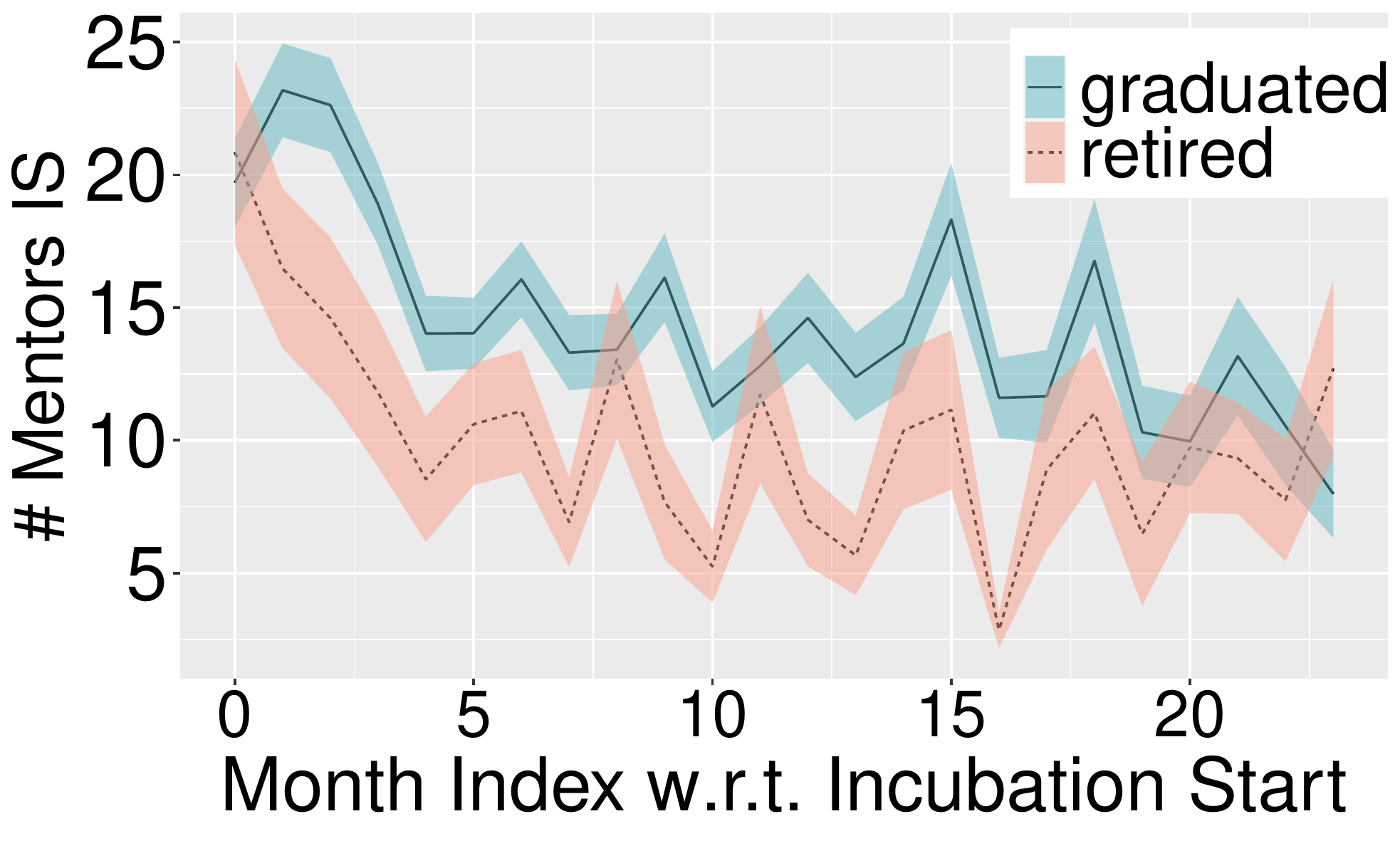}}
\subfigure[Num. IS from Committers]{
\label{IS_committers_curve}
\includegraphics[width=0.3\linewidth]{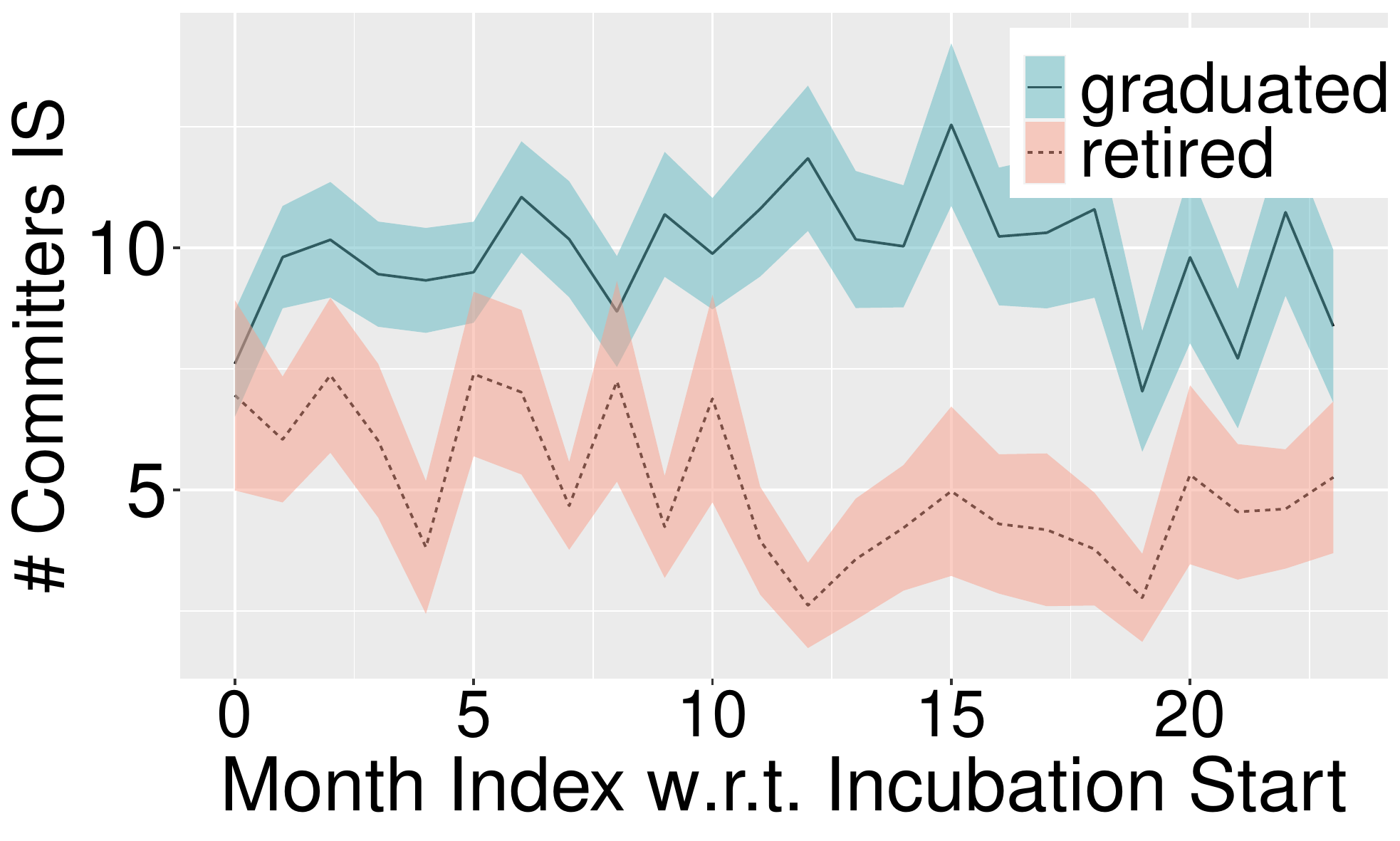}}
\subfigure[Num. IS from Contributors]{
\label{IS_contributors_curve}
\includegraphics[width=0.3\linewidth]{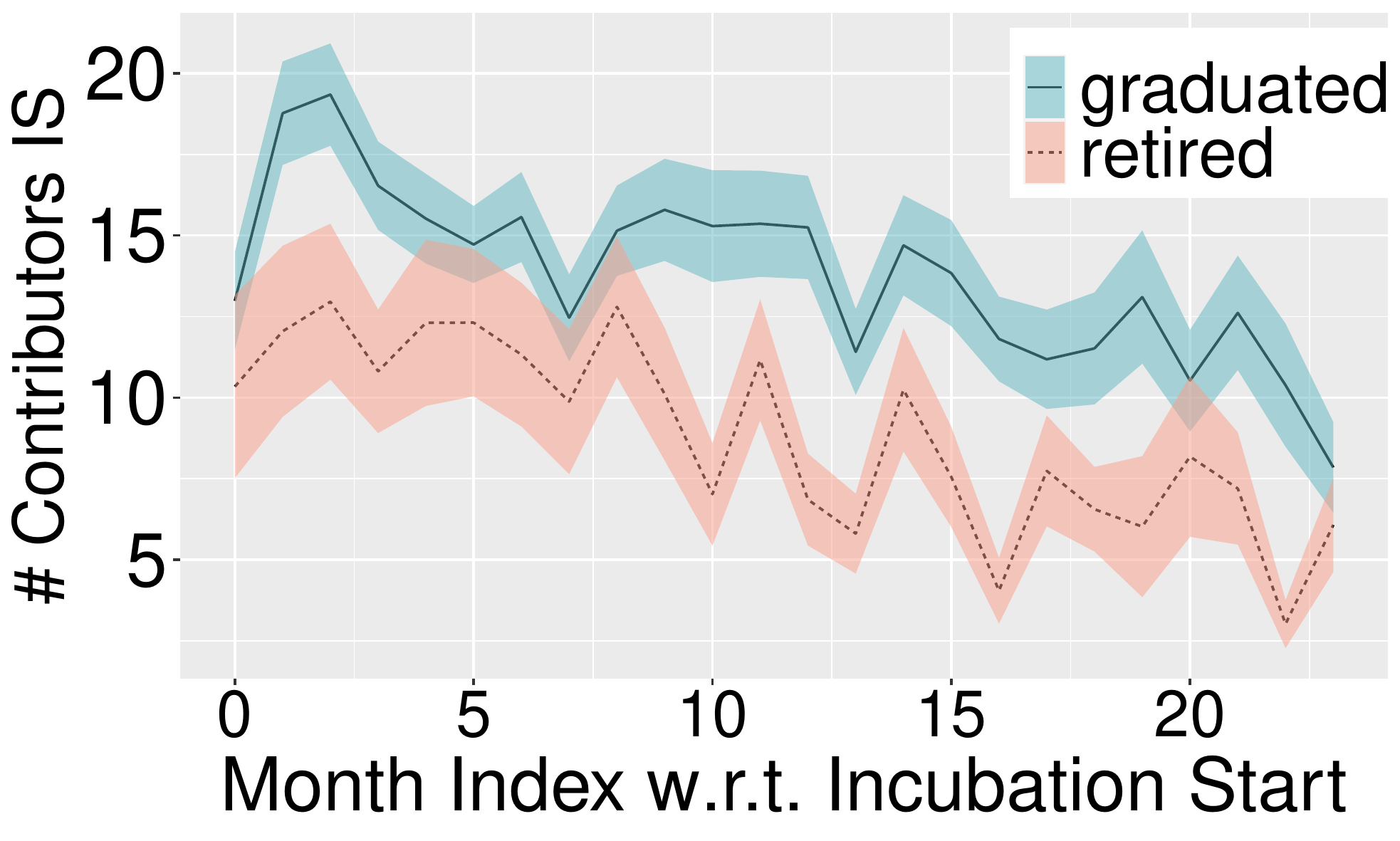}}
\subfigure[Developer Nodes in Social Network]{
\label{s_nodes_curve}
\includegraphics[width=0.3\linewidth]{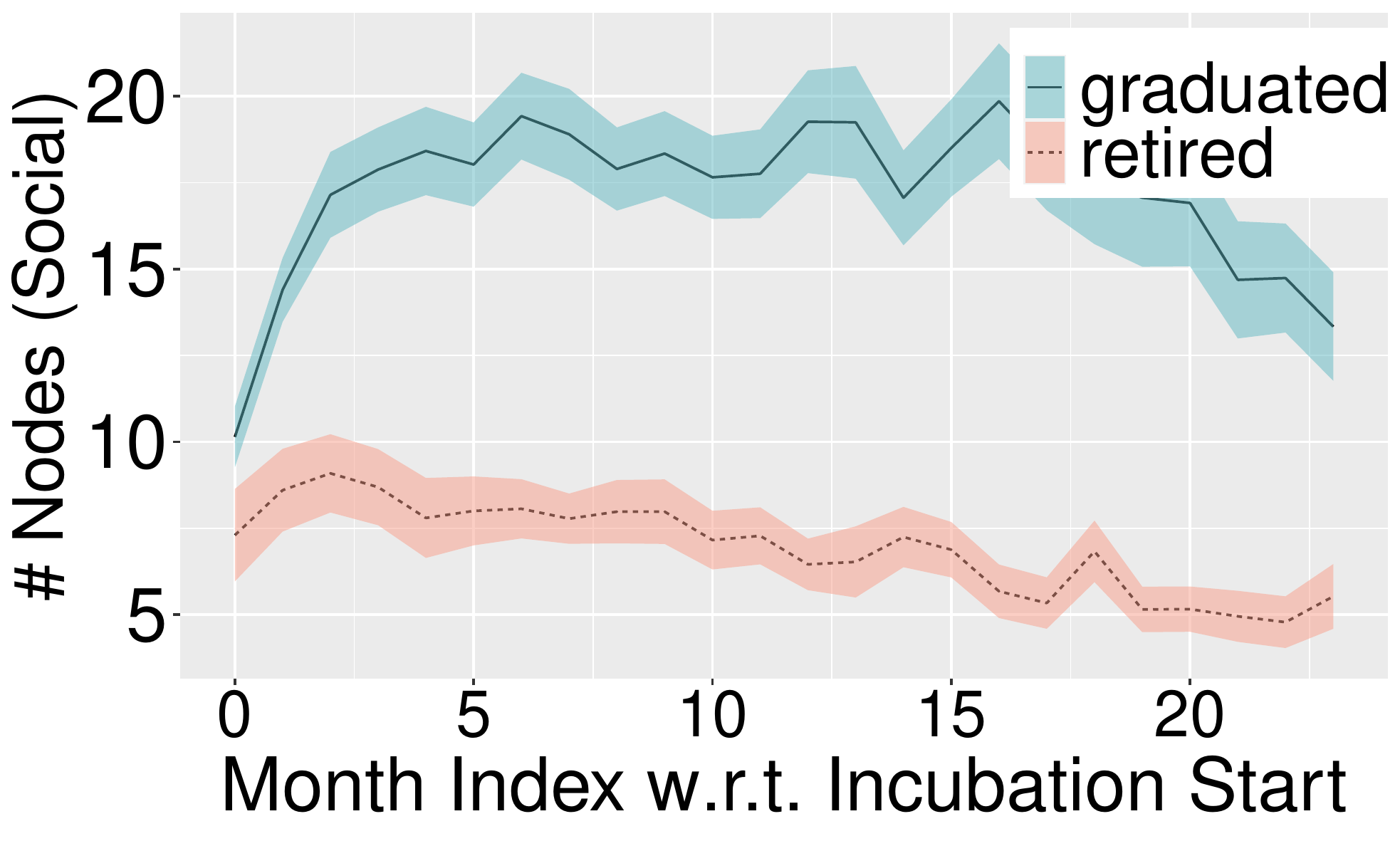}}
\subfigure[Developer Nodes in Tech Network]{
\label{t_dev_nodes_curve}
\includegraphics[width=0.3\linewidth]{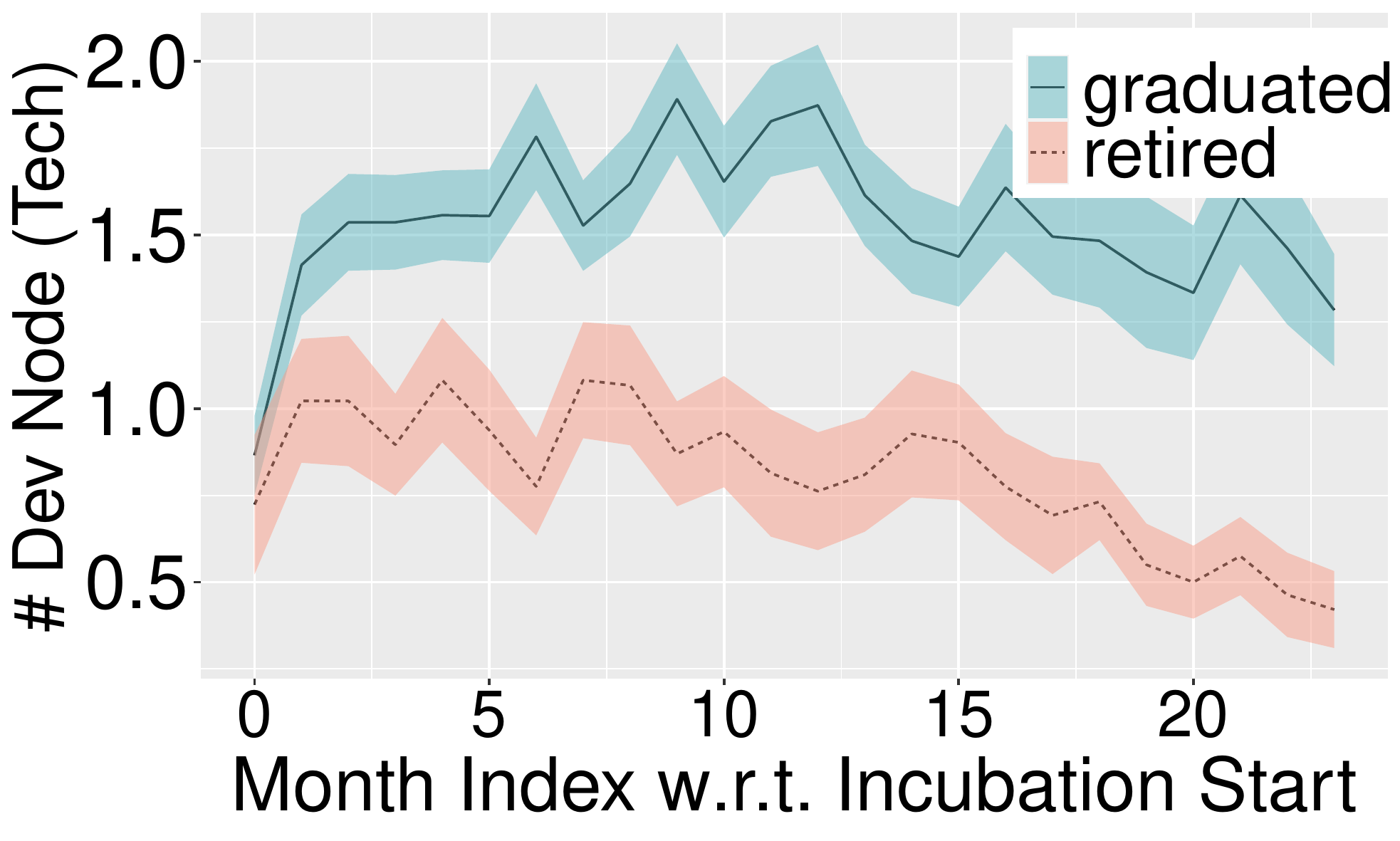}}
\subfigure[File Nodes in Tech Network]{
\label{t_file_nodes_curve}
\includegraphics[width=0.3\linewidth]{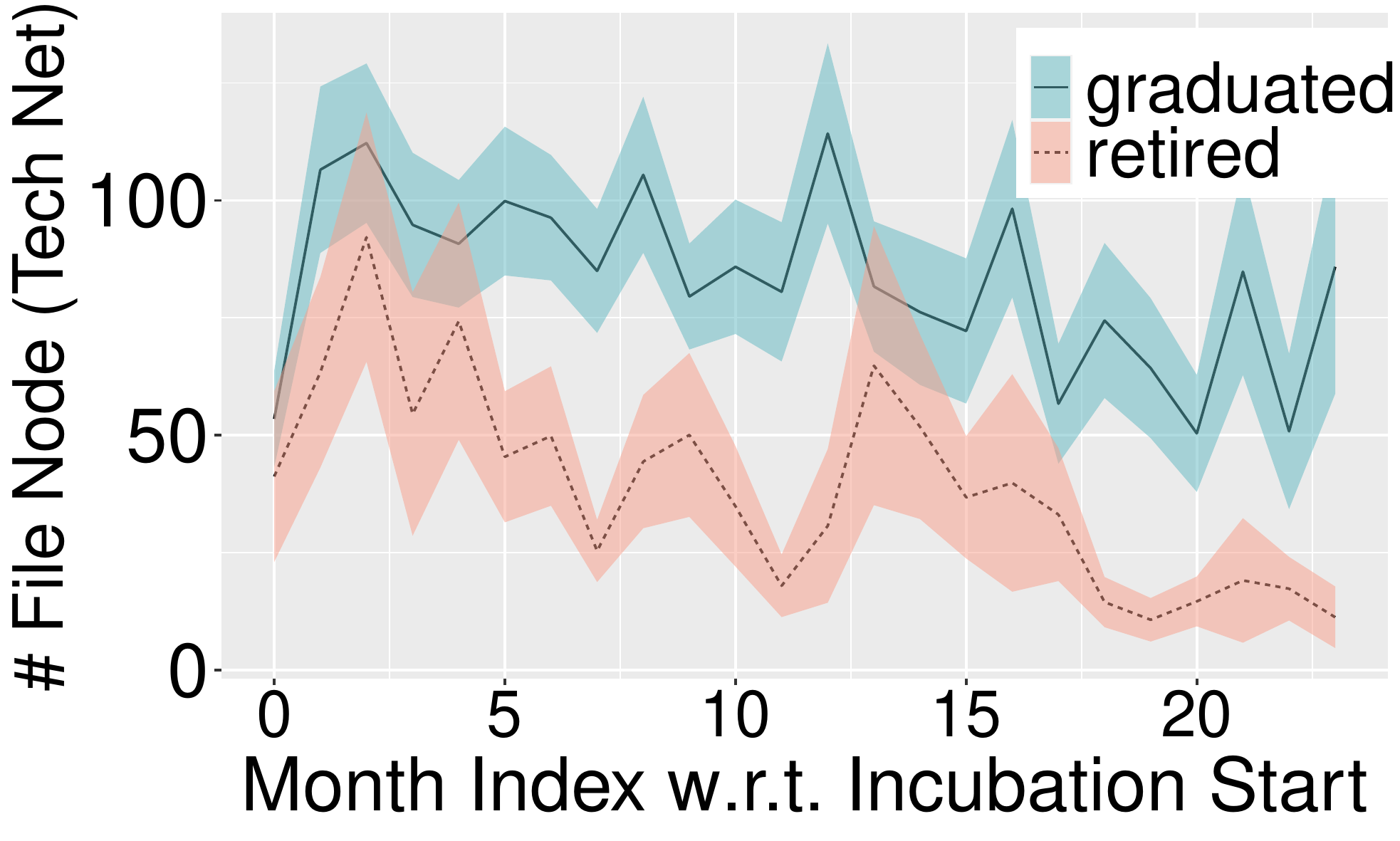}}
\caption{The averaged monthly IS and ST variables between graduated projects and retired projects. On the top are the IS measures; On the bottom are ST measures. Shades indicate one st. error away from the mean. Month index 0 indicates the incubation starting month.}
\label{fig:long}
\end{figure*}

We observe an increasing trend of Topic 1 `Progress Report' with a small seasonal effect, suggesting the projects are learning the `Apache Way' and more actively discussing their regular project reporting over time. And such seasonal effect is found to be more significant in the Topic 5 (`Report Review').
\textcolor{red}{Project releases, documentation, and software testing, are all connected to the number of people participating regularly. 
The retired projects are on average smaller than the graduated ones, which is the likely explanation for the differences. E.g., in Figure~\ref{t_file_nodes_curve}, we show that graduated projects, on average, have more source files than the retired projects.}
Moreover, we find that Topic 9, `license policy', is more prominent in the earlier stages of incubation (e.g., months 1-7), which makes sense in that the shift from one OSS license to the license required by ASF is an important discussion that projects would want to address earlier on.

On the contrary, the longitudinal pattern of IS language related to software testing is relatively rare in the beginning project incubation. 
It suggests that in earlier stages of incubation, developers are more likely focused on the transition to the incubator and perhaps less on new code development and testing. 
On the other hand, such transitions were implemented in fast manner, with testing discussions increasing rapidly in incubation months 3, 4 and 5.

By comparing graduated and retired projects, we find that, Topic 10, `Routine work', to be the dominant topic for both types of projects, almost through all projects' incubation (i.e., remain high volume compared to other topics).
We also find that graduated projects tend to be more active on Topic 7 `Documentation' and Topic 3 `Project Release'.
Interestingly, on the other hand, mentorship related ISs (Topic 11) are found to be more active for the retired projects rather than graduated projects. One possible reason is that retired projects did seek help from their mentors when their projects were experiencing downturns, and further issuing project-wise statements.

\underline{Metric Evolution.} We continue by exploring the evolution of our metrics over time. 
Looking at the mentors ISs, shown in Figure~\ref{IS_mentors_curve}, we can see that even in the beginning of their incubation, mentors email a greater number of ISs to projects that eventually graduate compared to ones that eventually retire.

Next, we see that the number of ISs in mentor emails decline for both graduated projects and retired projects before month $5$, suggesting that ASFI mentor activity may decrease after incubating projects work through the first steps of the incubation process.

Then, we visually identify an increasing trend of IS from mentors around month $6$ for graduated while $5$ for retired projects. 
One possible reason is the fact that mentors start helping projects when they are experiencing difficulties or downturns.
It is consistent with ASF mentorship that during the early stage of the incubation, developers are required to make institutional-related decisions, e.g., voting for reports, discussing the ASF required licensing, and the community-related issues, and it is in these kinds of areas where mentors come to help.

On the Socio-Technical networks side, shown in Figure~\ref{s_nodes_curve}, for the first 6 months, we can see the graduated projects have a clear increasing trend of the number of nodes in social networks, while retired projects seems to be constant. 
We can see a slight decrease around month 10 to month 12 for both types of projects, suggesting 10 months might be a good timing for mentors to intervene/motivate their projects, if they are experiencing some difficulties.

\begin{tcolorbox}[colframe=browna]
\textbf{RQ$_{\textcolor{myred}{2}}$ Summary:} We identify socio-technical and institutional signatures of OSS project evolution, and evidence that it differs between graduated and retired projects, and that these patterns can even be distinguished by institutional heuristic topics.
%On the institutional side, graduated projects receive more institutional guidance from their mentors over the course of incubation compared to retired projects. 
On the institutional side, both graduated and retired projects have more stable institutional topics during their first 3 months. 
On the Socio-Technical network side, graduated projects keep attracting community over their first 6 months, while retired projects are unstable during their first 3 months.
\end{tcolorbox}

%\note{Our original RQ2 in the RQs.txt file states this: RQ2: Is OSS project evolution toward sustainability readily observable through the dual lenses of institutional and socio-technical analysis? And how do such temporal patterns differ? I suggest we revise the Summary Box to this to tighten it up: "RQ2 Summary. Earlier we asked: Is OSS project evolution toward sustainability readily observable through the dual lenses of institutional and socio-technical analysis? And how do such temporal patterns differ? In this section we identified institutional and socio-technical signatures of OSS project evolution, and evidence that longitudinal patterns differ between graduated and retired projects, and that these patterns can even be distinguished by institutional heuristic topics.}

\subsection{Case Study: Association Between Institutional Governance and Organizational Structure}
To communicate concretely how the institutional and socio-technical dimensions interact within ASFI ecosystem, we showcase four diverse instances of their mutual interrelationship. 
%These cases illustrate how institutional discussion elicits and coordinates contributions to the code, while problems and events on the socio-technical side motivate institutional discussions.

\underline{Case A.} In July 2011, the \textit{HCatalog} project announced a vote for its first \textcolor{red}{Release Candidate (RC)}, the first officially distributed version of its code~\footnote{\url{https://lists.apache.org/thread/p88lpxn3gtprc11xw20jbnrmp3f8fmpw}}.
Because a project's RC's reflect on the whole ASF, they require approval from the foundation after project contributors have given their approval. 
In preparation for the first vote, developers double-checked the installation process and reported missing files and features. 
This drove contributions to the code and documentation, e.g., release notes were added after being reported missing. 
The contributors then cast their votes. With four votes, the product was approved and a proposal was forwarded to Apache Incubator leadership for approval.

\underline{Case B.} In December 2010, an independent developer emailed the \textit{Jena} project community to share their idea for a new feature, and asking how to proceed toward contributing it~\footnote{\url{https://lists.apache.org/thread/8dgbvjkwosbvxg6oj6ptyssn0m4rdto1}}.
Their query includes policy questions, such as whether they must obtain an Individual Contributor License Agreement (ICLA). 
A developer responds that the policy does not require an ICLA for the type of smaller contribution that the volunteer is proposing. 
The developer then guides the volunteer through established project processes for contributing to the code, including what mailing lists to use and how to submit their feature as a patch. 

\underline{Case C.} In December 2016, a developer in the \textit{Airflow} project community raised concerns over the integration testing infrastructure offered by Apache, citing unnecessary obstacles it imposes on volunteer contributors\footnote{\url{https://lists.apache.org/thread/twsg2h72d2025jc6nc1ltzgbmwoxh7cx}}. 
The developer offers their own resources as an alternative, with the caveats that they will administer it and control access. 
This triggers a discussion on the technical merits of the developer's concerns, and a policy discussion as to whether ASF permits the use of unofficial alternative infrastructure options. 
Several developers conclude that a transition is technically advisable and institutionally sound, and the community transitions to the alternative integration testing framework. 

\underline{Case D.} In September 2015, \textcolor{red}{the \textit{Kalumet} project received a proposal that it be retired from ASFI} after its code had been languishing for  several months. 
Contributors agreed upon retirement almost unanimously. 
One contributor, identifying  features of the project that could be of use to other ASF and ASFI projects, suggests distributing key parts of it functionality to other active projects.
The retirement vote is ultimately followed by developer effort distributing \textit{Kalumet}'s assets.

These cases illustrate how institution-side policy discussion and sociotechnical-side project contributions interact, with developments on the artifact motivating policy discussions, and policy constraints steering developer effort. 
With longitudinal data on both institutional and socio-technical variables, we now transition to a quantitative investigation of these relationships.

\subsection{RQ$_{\textcolor{myred}{3}}$: Are periods of increased Institutional Statements frequency followed by changes in the project organizational structure, and vice-versa?}

\begin{figure}[tbp]
\centering
\subfigure[Granger Test for Graduated Projects ($p < .001$)]{
\label{s_mean_degree_social_net}
\includegraphics[width=0.8\linewidth]{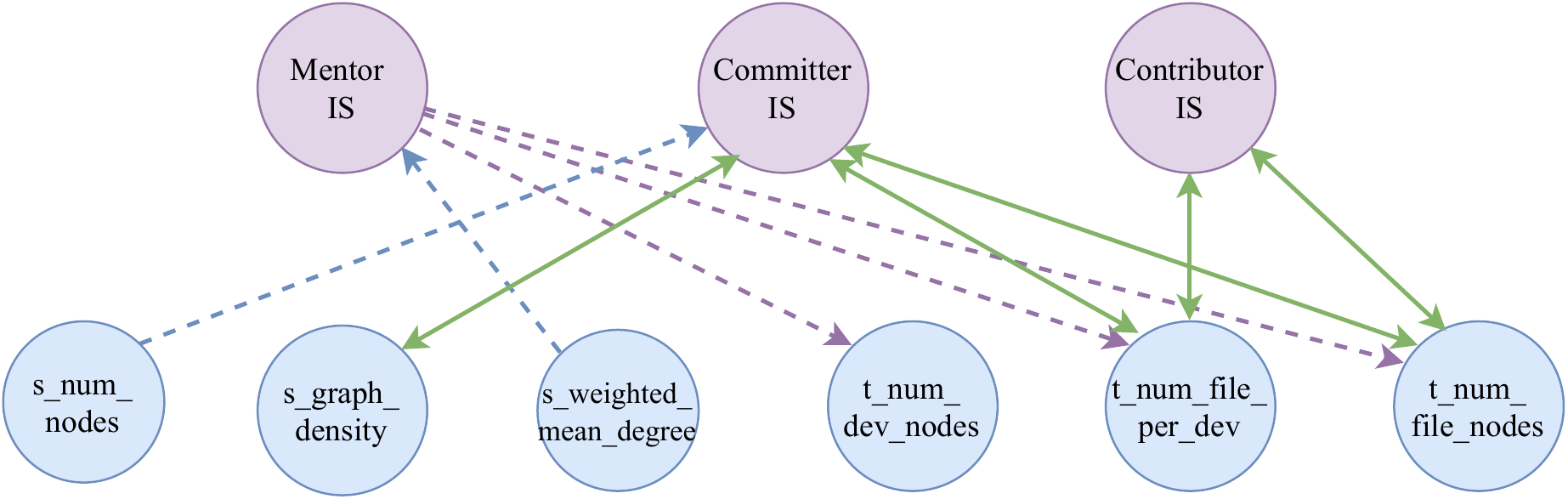}}
\subfigure[Granger Test for Retired Projects ($p < .001$)]{
\label{num_file_node_technical_net}
\includegraphics[width=0.8\linewidth]{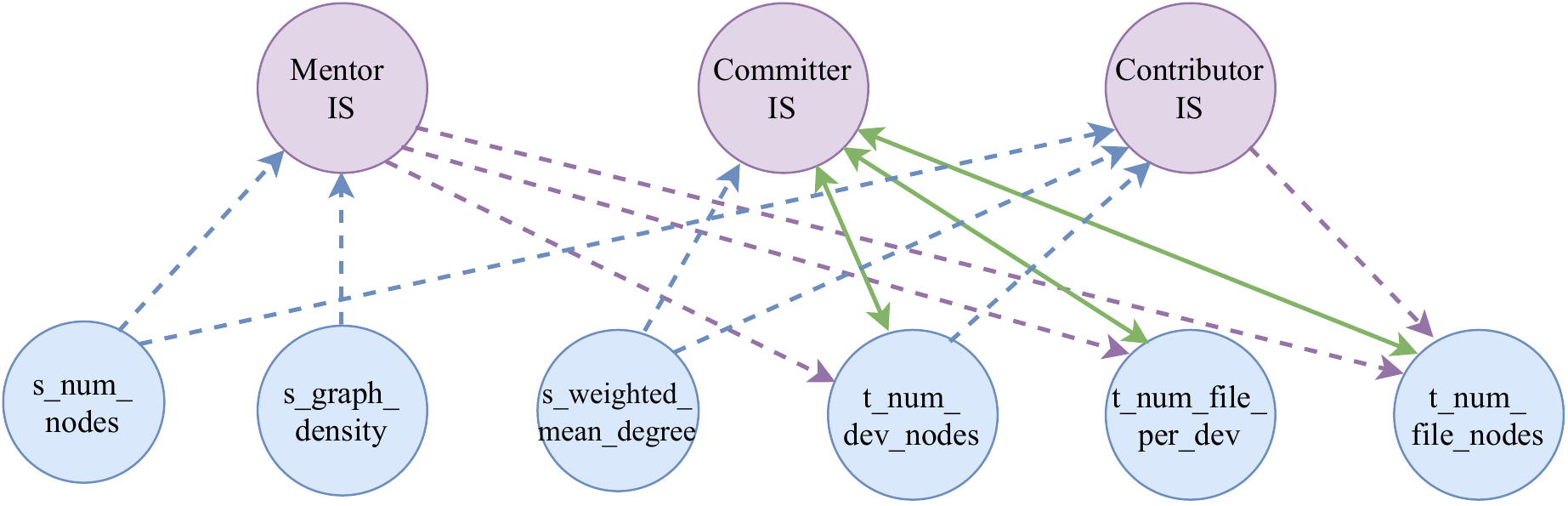}}
\caption{The Granger Causality between Institutional Statements and Socio-Technical networks. The blue/purple directed links indicate Granger causality from ST/IS measures, respectively. A green bi-directional link indicates that there is two-way significant temporal relationship (p-value < 0.01). Graduated projects seem to have fewer links from ST variables to IS variables, suggesting a more unidirectional flow from institutional to sociotechnical changes in successful projects. }
\label{fig:granger}
\end{figure}

In the previous RQs, we conducted exploratory and qualitative studies of the IS extraction technology described in Section 4.3, and of IS and socio-technical variable changes over time.
In this section, we investigate the temporal relationship between our measures of IS discussion and ST networks, as OSS projects progress on their incubation trajectories.
As predicted by contingency theory, our hypothesis is that during project evolution, developers and mentors must make time for decisions related to their organizational structure, contingent on ASF required institutional arrangements and governance. 
That is, incubating projects change their organizational structure based on the institutional norms and rules being discussed, as required of them as potential new member of the ASF community.
And vice versa, organizational changes can incite followup discussion about institutional processes.
To test for RQ$_{\textcolor{myred}{3}}$, here we use the pair-wise Granger causality test with lagged order of $2$. 
We run the test for all pairs between the Institutional Statements and socio-technical sets of variables, resulting in 36 separate tests for the graduated projects set and 36 for the retired ones.
We adjust the $p$-values for multiple hypothesis testing since we test 72 hypotheses, using the Benjamini-Hoschberg procedure~\cite{ferreira2006benjamini}. 
We only consider as significant p-values $< 0.01$.
The results are summarized in Figure~\ref{fig:granger}, where a directed edge from $X$ to $Y$ indicates that $X$ Granger-causes $Y$. 
Recall that these are lagged relationships, over time, and that the roles of people may change over time, e.g., a contributor this month may become a committer next month.
Also, as discussed in Section 4.6, Granger is not a complete test of causality, but does yield and effect and its directionality, although without effect size or sign.

%I.e., there exists significant difference between using historical data from only variable $Y$ and using both historical data from $X$ and $Y$, to predict the current value of $Y$.  

%\sout{\underline{Discussion.}} 
We observe a large number, 31 (out of 72 total), of Granger-causality relationship between the institutional statements frequency set and the socio-technical network measures. 
15 of them are from the graduated set and 16 from the retired set.
8 of the relationships are shared between the sets.
We conclude that there is a significant Granger-causality between changes in institutional governance discussions and organizational structure of the projects.
We note 8 bidirectional relationships, the remaining 15 are unidirectional\footnote{Bidirectional causality indicates a feedback of some sort. E.g., supply causes demand and demand, in turn, causes supply; a conversation with a child is another example, where a question causes a response which in turn incites another questions.}.

We look at graduated projects first.
Interestingly,  Figure~\ref{fig:granger}, top, shows that the number of ISs from mentors, committers and contributors have effects on the technical network, and vice-versa for the latter two.
Namely, they all Granger-cause network changes on the technical end, i.e., on developer file productivity  (\texttt{t\_num\_files\_per\_dev}), and total number of files changed (\texttt
{t\_num\_file\_nodes}) variables.
Mentor IS, additionally, Granger-cause changes to number of developer (\texttt
{t\_num\_dev\_nodes}).
This is consistent with ASFI expectations that a mentor's emails provide advice and engage people, and conversely, that a drop in engagement may elicit a mentor engagement.

\begin{comment}
One explanation for this is that contributors through their ISs may induce subsequent coding actions by themselves or others.
\end{comment}

Mentors usually do not code, which is presumably why they Granger-cause but do not appear in feedback relationships with any of the technical network variables.

Notably absent, however, are links from mentor and contributor ISs into social network variables.
Only committer ISs (bidirectionally) Granger-cause changes in the social network density, which, perhaps, simply indicates that ISs from committers induce substantial  traffic in the social network, which in turn get committers to discuss policy and rules issues.
%Similarly, change in social numbers causing commiter IS changes possibly indicates that members engage in policy discussions about file changes. 
\textcolor{red}{We have observed situations where mentors are likely to interrupt the projects when the projects become less active (either socially or technically)\footnote{\textcolor{red}{An example of mentor interrupting project \emph{warble}: \url{https://lists.apache.org/thread/x6h8pzhmfwtyy354ml1xm9sylq4y5r7l}}}}.
On the other hand, it could also be that a mentor is reacting to some particular broader discussion among developers, e.g., one on a monthly report.

Together, the above tell a story of the importance to the technical networks of changes in any IS variable. 
Surprisingly, mentor IS changes are not as consequential to the social network, seemingly at odds with the ASF community-first goals.
Thus, there may be room to enhance community engagement with mentors and vice-versa.

%Looking at the retired projects' causality network, Figure~\ref{fig:granger}, bottom, we see a number of differences compared to the graduated one, top.
\begin{tcolorbox}[colframe=browna]
\textbf{RQ$_{\textcolor{myred}{3}}$ Summary:}
In both graduated and retired projects, there are no inputs from the IS into the social network variables, even though there are IS inputs into all technical network variables.
Retired projects exhibit less bidirectionality between ST and IS variables.
Finally, and interestingly, among retired projects there are causal inputs into contributor ISs from both the social and technical variables. This is not the case for the graduated projects.
%Overall, graduated projects have fewer links from socio-technical variables in the institutional direction, suggesting that successfully graduated projects manifest a more unidirectional flow from institutional communication to sociotechnical change.
\end{tcolorbox}
%\end{document}

\section{Discussion}
%\documentclass[main.tex]{subfiles}
%\begin{document}

\textcolor{red}{In this study, we use individual institutional prescriptions, Institutional Statements (ISs), and Socio-Technical network structure to reason about OSS project sustainability. We measure project sustainability by whether an OSS project graduates or retires from the ASF Incubator.}

\textcolor{red}{OSS projects are a form of digital public goods which, like other public goods (e.g., water, forest, marine, etc.), can be subject to degradation due to over-harvesting, e.g., in the form of free-riders who take advantages of OSS but do not contribute to the required resources for development and maintenance of the software.}
\textcolor{red}{Ostrom's work illuminated the fact that many communities avoid the dreaded `Tragedy of the Commons', and other collective action problems, through the hard work of designing and implementing self-governing institutions.}
\textcolor{red}{In that context, the ASF is a nonprofit foundation that, through its incubation program, encourages nascent OSS projects to follow some ASF-guided operational-level rules or policies around their self-governance. The OSS projects that join the ASF incubator trade some of the freedom of unlimited institutional choice in exchange for Incubator resources that increase their chances of enduring the collective action problems that characterize OSS development ~\cite{schweik2007tragedy}, and becoming sustainable in the long run.}

\textcolor{red}{We find that in the ASF Incubator, the amount of institutional statements and levels of socio-technical variables are associated with OSS sustainability. 
In particular, in RQ$_{\textcolor{myred}{1}}$, we show that graduated projects are larger, and in them there are more ISs by all three types of participants: committers, contributors, and project mentors, than in the retired projects. This, presumably, is indicative of more active or intentional self-governance.}
\textcolor{red}{In  theoretical and empirical work on commons governance, it is well documented that getting self-governing institutions "right" is hard work and takes time and effort ~\cite{ostrom2009understanding}. This may imply our finding, that in graduated projects more ISs are expressed  compared to retired projects. This is consistent with a narrative that participants in graduated projects debate and work harder on their project's operational-level institutional design.}

\textcolor{red}{Socio-Technical Systems (STS) theory posits that the social and technical components of a system influence each other. Recent work has shown that ASFI retired and graduate projects have sufficiently different socio-technical structures~\cite{yin2021sustainability}. In light of that, the results in RQ$_{\textcolor{myred}{2}}$, showing that the topics of institutional-relevance in the graduated projects differ from those of the retired projects, are consistent with STS theory.
Moreover, our methods point to a way to study possible interventions in underperforming projects, based on the topics discussed in their institutional statements.
However, even among graduated projects there is still diversity in the institutional statements. Thus, as predicted by contingency theory, as well as Ostrom's theory of institutional diversity~\cite{OstromJanssenAnderies2007}, a one-size-fits-all solution to a successful trajectory toward sustainability is not likely. Instead, future work should focus on gathering larger corpora of data, to be able to resolve the individual or small-group differences in successful projects.}

\textcolor{red}{Finally, our framework allowed us to combine the IS and STS structures and study them together over time. With it, in RQ$_{\textcolor{myred}{3}}$, we find  two-way, causal correlations between socio-technical variables and ISs over time, arguably indicating that OSS project socio-technical structure and their governance structure evolve together, as a coupled system.}

\textcolor{red}{Moreover, we find evidence supportive of a prediction of commons governance theory (Sect. 2.1), that the ASF Incubator governance infrastructure can effect collective action in the projects. 
Specifically, we find in the answer to RQ$_{\textcolor{myred}{3}}$ that mentors react to socio-technical actions by participants in the projects, and vice-versa, though the feedback and information flow is different between graduated and retired projects.
This again opens entry points and targets for interventions, whereby underperforming projects could be leaned on via rules or advice in order to adjust their trajectories.
}

%\end{document}

%\documentclass[main.tex]{subfiles}
%\begin{document}
\section{Threats to Validity}
First, our commit and email data is from only hundreds of projects ASF incubator projects. 
Thus, generalizing the implications beyond ASF, or even beyond the ASF Incubator projects carries potential risks, \textcolor{red}{for example, OSS projects in other incubator programs may not have mentors.}
Expanding the dataset beyond ASF incubator, e.g., with additional projects from other OSS incubator program could lower this risk. 
Second, we do not consider communication channels other than the ASF mailing lists, e.g., in-person meetings, website documentation, private emails, etc. 
However, ASF mandates the use of the public mailing lists for most project discussion, a policy that ensures particularly low risk of missing institutional or socio-technical information~\footnote{The Apache Way: \url{http://theapacheway.com/on-list/}}.
Annotations of the Institutional Statements (IS) can be biased by individual annotators, while we gave the annotators sufficient training and a reference documentation which lowers the risk. 
We expect the performance of the classifier as we increase the size of the training set and better incorporate contextual information, and we plan in the future to distinguish types of ISs.
In OSS projects, developers may use their different emails or aliases, which in turn complicates the identification of distinct developers, while assigning and insisting using an unique apache.org domain email address reduces such risks~\footnote{ASF committer emails: \url{https://infra.apache.org/committer-email.html}}.

\textcolor{red}{Finally, as noted in Sect. 4, there are likely cases where OSS projects that
have retired from the ASF Incubator program still go on to become sustained over time. In these
instances, some OSS projects entering the ASFI may simply not be a good fit to the ASF culture and
institutional requirements or policies and ultimately retire as a result. In this paper we explicitly
use graduation as a measure of sustainability given that this is an ultimate goal of the ASFI – to
create projects that can indeed be sustainable. But we want to recognize the point that few retired
projects still could become sustainable following a different path than association with ASF.}

\section{Conclusion}
Understanding why OSS projects cannot meet the expectations of nonprofit foundations may help others improve their individual practice, organizational management, and institutional structure.
More importantly, understanding the relationship between institutional design and socio-technical aspects in OSS can bring insights into the potential sustainability of such projects.  
Here we showed that quantitative network science features can capture the organizational structure of how developers collaborate and communicate through the artifacts they create.
Moreover, we show that institutional statements in OSS projects can be predicted by state-of-the-art deep learner methods. 
Combining the two perspectives, socio-technical measures and institutional analysis, we leverage the unique affordances of the Apache Software Foundation's OSS Incubator project to extend modeling of OSS project sustainability, leveraging a novel longitudinal dataset, a vast text and log corpus, and extrinsic labels for the success and failure of project sustainability.
Future work is needed to offer validation of this or similar strategies experimentally.
%\end{document}

\bibliographystyle{acm}
\bibliography{ref}

\end{document}